\documentclass[a4paper]{panl}
\usepackage{cite}
\usepackage{wrapfig}
\usepackage{graphicx}
\usepackage{amssymb}
\usepackage{amsfonts}
\usepackage{amsmath}
\usepackage{longtable}
\usepackage{rotating}
\usepackage{lscape}
\usepackage{epsfig}
\usepackage{multirow}
\originalTeX
\begin{document}

\title{
MODEL FOR CLASSICAL ELECTRON\\
WITH FINITE MASS AND ACTION\\
}
\maketitle
\authors{S.\,I. Manaenkov$^{a,}$
\footnote{E-mail: manaenkov$_-$si@pnpi.nrcki.ru}}
\from{$^{a}$\,National Research Centre "Kurchatov Institute" Petersburg Nuclear Physics Institute,\\
Gatchina, Leningrad region, 188300, Russia}

\begin{abstract}

It is shown in the tetrad representation that
there are Reissner--Nordstr\"om solutions with a finite action and  total inertial mass
equal to the gravitational mass of the considered system. These solutions describe systems of electromagnetic and gravitational fields
without any admixture of massive point-charges. The stress tensor for this solutions is shown to be identically zero. This means that there is no need in additional 
nonelectromagnetic surface-tension, existing in the Lorentz electron model, preventing the system disintegration.
The hypothesis
that gravitation can play
a crucial role in the structure of elementary particles is discussed.
\end{abstract}

\vspace*{6pt}
\noindent


\section*{Introduction}
\label{sect-1}
 As is well known the ratio of the electrostatic repulsion between two electrons is larger than their gravitational attraction by a 
factor of $4.2\cdot  10^{42}$. At first sight, this means that gravitational interaction cannot play important role in the elementary 
particle interaction and in their structure. As will be argued this point of view can probably be wrong. 

The main concept of classical electrodynamics (see for instance \cite{JDJ,LL}) assumes the existence of charged massive particles  and the 
electromagnetic field responsible for their electromagnetic interaction. The principle difficulties
 of this concept are  well known: if the fundamental charged particles 
  are point-like, then the energy 
of the  electromagnetic field is infinite even for one particle.
If the particle is considered as a drop of charged liquid, as in the Lorentz model of an electron\cite{LOR}, having a 
finite range in the  three-dimensional space, the problem of the nature of non-electromagnetic attractive forces  preventing  the drop 
disintegration by the electrostatic forces arises immediately. These problems are not solved up to now. 

In quantum electrodynamics (QED),  electron-positron field  is  a bispinor field of point-like particles, which electromagnetic current 
is a source of the  quantized  electromagnetic field. The lagrangian  of QED predicts an infinite mass  and infinite electric charge  of 
the electron and positron  (see for instance \cite{RPF,AB,Bog}). In order to give physical meaning to divergent integrals and to get the experimentally observed 
cross sections, the procedure of the mass and  charge renormalization was invented. But this infinite renormalization is realized by an 
introduction into the initial QED lagrangian of contra-terms having infinite coupling constants \cite{Bog}.

In the present paper, another approach will be discussed in which the classical electron is considered as the system of electromagnetic and the gravitational 
fields described by the Reissner--Nordstr\"om (RN) solution of both the Maxwell and Einstein equations. 
It will be shown that fields are localized in a  very small three-dimensional-space region (about $10^{-34}$ cm). To avoid any misunderstanding, we would like to 
stress 
that the main results are obtained in the framework of classical physics. Only in Secs. \ref{sect-10} and \ref{sect-11}, quantum effects will be shortly discussed.
It will be demonstrated that in spite of the singularities  of the electromagnetic 
and gravitational fields, the total lagrangian density is integrable function and the action is finite in the tetrad representation for any values  of the parameters 
$e$ and $m$ of the RN solution denoting the electric charge and the gravitational mass of the system, respectively. As will be shown  if the parameters
 $e$ and $m$ obey the   relation, considered in the present paper, the solution  corresponds to a finite total inertial mass of  the system 
that is equal to its gravitational mass. This means that the solution is in accordance with the equivalence principle.  
It is this solution with the charge equal to the experimental electron charge which will be quoted as  ``the classical  electron".
Note that the total mass  is just the  mass of the electromagnetic   and gravitational fields. This means that there is no need in any additional  fundamental entities 
such as charged point-like particles (electron-positron field) and the only  existing fields are the electromagnetic and gravitational fields.

The paper is organized as follows. In Sec.~\ref{sect-2}, the RN solution is presented in the spherical coordinates and the main notations are explained,
while the RN solution in the uniform coordinates is given in Sec.~\ref{sect-3}. The tetrads, formulas for their covariant derivatives, the tensor $\gamma_{ikl}$ and
vector
$\Phi_j$ expressed in terms of the tetrads are considered in Sec.~\ref{sect-4}, besides the detailed formulas are presented in Appendix in Subsections 
\ref{appendix-1},
\ref{appendix-2}, and \ref{appendix-3}. Formulas for the total lagrangian density and lagrangian for the RN solution in the tetrad representation are obtained in 
Sec.~\ref{sect-5}.  The Lagrange equations for   
the tetrads are discussed in Sec.~\ref{sect-6}. It is checked in Subsec.~\ref{appendix-4} of Appendix that the used tetrads obey the Lagrange equations.
It is proved in Sec.~\ref{sect-7} that the total energy-momentum pseudotensor density is integrable function for the classical electron. Therefore  its total inertial 
mass 
is shown to be finite. It turns out that the stress tensor density is identically zero for the classical electron. 
It is shown in Sec.~\ref{sect-8} that according to the equivalence principle there is no need in any point-like particles
with some bare mass that are usually named electrons.
The electrical charge distribution is considered in 
Sec.~\ref{sect-9}. 
Hypotheses concerning the quantum generalization of the obtained results are discussed in Sec.~\ref{sect-10}. The most important results are summarized in 
Sec.~\ref{sect-11}.

\section{Reissner--Nordstr\"om solution}
\label{sect-2}
Let us consider the spherical coordinates $x^1=r$, $x^2=\theta$, $x^3=\varphi$, and $x^0=x_0=ct$, 
 where $r$ denotes a radius, $\theta$, $\varphi$ 
are polar and azimuthal angles, $c$ is the velocity of light in empty space and $t$ denotes time. 
The stationary solution of the Maxwell and Einstein equations, depending only on  $r$, was found independently
by H.~Reissner, H.~Weyl, G.~Nordstr\"om, and G.B.~Jeffery \cite{HR,HW,GN,Jeff}, nevertheless it is called ususally the Reissner--Nordstr\"om solution. 
From here on, it will be referred to as the RN solution.
For this solution, the metric tensor can be chosen diagonal and its nonzero covariant and contravariant  components are
\begin{eqnarray}
g_{00}=1/g^{00}=-g^{11}=-1/g_{11}=\Lambda,
\label{g00-fin}\\
\Lambda=1-\frac{2 k m}{c^2r}+\frac{k e^2}{c^4r^2},
\label{grr-fin}\\
g_{22}=1/g^{22}=-r^2,
\label{g22}\\
g_{33}=1/g^{33}=-r^2\sin^2\theta.
\label{g33}
\end{eqnarray}
Here, $k=6.67 \cdot 10^{-8}$ cm$^3$ $\cdot$ g$^{-1}$ $\cdot$ s$^{-2}$ is the gravitational constant, 
$e$ and $m$ are respectively the electric charge and  mass of the considered 
system.  This solution is used in cosmology to describe black holes
for which
\begin{eqnarray}
m> \sqrt{\frac{e^2}{k}}.
\label{black-hole}
\end{eqnarray}   
In this work, we start our consideration with the  cases when the inverse condition 
\begin{eqnarray}
m< \sqrt{\frac{e^2}{k}}
\label{cond-1}
\end{eqnarray}
is fulfilled since it is this condition which is valid for real leptons and quarks. We shall name such a system ``the point-charge" which could correspond to the 
electron if 
$e=-4.80 \cdot 10^{-10}$ esu. We shall distinguish ``the real electron" with the experimental
mass $m=9.11 \cdot 10^{-28}$ g ($0.511$ MeV$/c^2$) and ``the classical electron"
which properties will be explained later.

It is convenient to rewrite Eq.~$(\ref{grr-fin})$ for $\Lambda$ in the form
\begin{eqnarray}
\Lambda=1-\frac{r_g}{r}+\frac{r_e^2}{r^2},
\label{lamb-1}
\end{eqnarray}
where
\begin{eqnarray}
r_g=\frac{2km}{c^2},
\label{rg}\\
r^2_e =\frac{k e^2}{c^4}
\label{re}
\end{eqnarray}  
with $r_g$ being the Schwarzschild radius \cite{Schw}. Using 
values $c=2.998\cdot 10^{10}$ cm/s and $e$, $m$ for the real electron
one gets $r_g=1.35 \cdot 10^{-55}$ cm, 
$r_e=1.38 \cdot 10^{-34}$ cm.
 As follows from these numerical estimates $r_e \gg r_g$ for the real electrons.

It is obvious from Eqs.~(\ref{g00-fin}) and (\ref{lamb-1}) that the element $g^{00}$ at $r\geq 0$
has no singularity and is positive when $r^2_{0}>0$, 
where $r^2_0$ is given by
\begin{eqnarray}
r^2_0=r^2_e -r_g^2/4\equiv r^2_e\Bigl (1-\frac{k m^2}{e^2} \Bigr ).
\label{r0}
\end{eqnarray}
Since according to the numerical estimates presented after Eq.~$(\ref{re})$ $r_g/r_e \ll 1$ we have from Eq.~$(\ref{r0})$ $r_0 \approx r_e$ with
a very high precision. We conclude also from the right-hand side of   Eq.~(\ref{r0}) that the requirement $r^2_{0}>0$ is equivalent to the  
condition~$(\ref{cond-1})$.
We  get from Eq.~$(\ref{cond-1})$  the numerical estimate $m<1.86 \cdot 10^{-6}$~g. This means that condition
$(\ref{cond-1})$ is fulfilled not  only for  real electrons but for muons, $\tau$-leptons, and quarks. 

The tensor of the electromagnetic field $F^{ik}$ obeys the Maxwell equations in the 
gravitational field \cite{LL,SW}:
\begin{equation}
\frac{1}{\sqrt{-g}}\frac{\partial}{\partial x^k} \Bigl \{ \sqrt{-g} F^{ik}\Bigr \}=-\frac{4 \pi}{c}J^i
\label{maxwell}
\end{equation}
with zero electromagnetic current $J^i= 0$ for all $r$ except  $r=0$.
Summing over any pair of identical covariant and contravariant Latin indexes  is assumed in Eq.~$(\ref{maxwell})$ and in all below formulas
except formulas in Appendix, where all sums have the symbol $\Sigma$. All Latin indexes can be equal to 0, 1, 2, 3, while the Greek 
indexes can be 
equal to 1, 2, 3. Summing over any pair of identical  Greek indexes is also assumed everywhere except Appendix. 
Here, $g$ is the determinant of the matrix $g_{ij}$.
 The solution of Eq.~$(\ref{maxwell})$ for the radial component of the electric field density $E_r$ is \cite{HR,HW,GN,Jeff}
\begin{equation}
F^{10}=-F^{01}=-F_{10}=F_{01}=E_r=\frac{e}{r^2},                 
\label{sol-F01}
\end{equation}
all the other components of $F^{ik}$ and $F_{ik}$ are zero. 

The formula for the energy-momentum tensor of the electromagnetic field reads \cite{LL,SW}
\begin{equation}
T^{i}_k=\frac{1}{4 \pi}\Bigl [ -F^{il}F_{kl}+\frac{1}{4}\delta^{i}_{k}F_{lm}F^{lm}\Bigr ],
\label{Tik} 
\end{equation}
where $\delta^{i}_{k}$ denotes the Kronecker symbol ($\delta^{0}_{0}=\delta^{1}_{1}=\delta^{2}_{2}=\delta^{3}_{3}=1$, 
all the other elements are zero). 
Using Eqs.~$(\ref{sol-F01})$ and $(\ref{Tik})$
one gets the formulas for the nonzero elements of $T^{i}_{k}$:
\begin{equation}
T^{0}_0=T^{1}_1=-T^{2}_2=-T^{3}_3=\frac{e^2}{8 \pi r^4}. 
\label{T-00}
\end{equation}
The metric tensor $g_{i k}$ and the energy-momentum tensor $T_{i k}$ 
obey the equations for the gravitational field established 
by Einstein \cite{Einst} and Hilbert \cite{Hilb}. 
\newpage
\section{Isotropic coordinates}
\label{sect-3}
In order to introduce the isotropic coordinates, we define the new radial variable $\rho$ with the relations
\begin{eqnarray}
r=\rho \mathcal{D}(\rho),
\label{r-rho}\\
\mathcal{D}(\rho)=1+\frac{r_g}{2 \rho }-\frac{r^2_0}{4 \rho ^2}\equiv \Bigl [1+\frac{r_g}{4 \rho } \Bigr ]^2 -\frac{r^2_e}{4 \rho ^2},
\label{def-D}
\end{eqnarray}
where $r_g$,  $r^2_e$, and $r^2_0$ are given by Eqs.~(\ref{rg}--\ref{r0}). Defining 
\begin{eqnarray}
\mathcal{N}(\rho)\equiv \frac {dr}{d\rho}=1+\frac{r^2_0}{4 \rho ^2},
\label{def-N}
\end{eqnarray} 
we get the formula for $\Lambda$
\begin{eqnarray}
\Lambda=\frac {\mathcal{N}^2}{\mathcal{D}^2},
\label{lambda-n-d}
\end{eqnarray}
and the following relation for the spacetime interval: 
\begin{eqnarray}
ds^2=\frac {\mathcal{N}^2}{\mathcal{D}^2}dx_0^2-\mathcal{D}^2\{d\rho^2+\rho^2(d\theta ^2+\sin ^2 \theta 
d\varphi^2)\}.
\label{def-ds2}
\end{eqnarray}
Let us introduce the pseudo-Euclidean coordinates $\rho^0 =\rho_0=x^0$, $\rho_x\equiv \rho^1$, $\rho_y\equiv \rho^2$, $\rho_z\equiv 
\rho^3$ 
(called the uniform coordinates) with the relations
\begin{eqnarray}
\rho_x=\rho \sin \theta \cos \varphi,\;
\rho_y=\rho \sin \theta \sin \varphi,\;
\rho_z=\rho \cos \theta.
\label{def-rhoxyz}
\end{eqnarray}
Now, the formula for the spacetime interval looks like
\begin{eqnarray}
\nonumber
ds^2= g_{ik}d\rho^id\rho^k= \\
=\frac {\mathcal{N}^2}{\mathcal{D}^2}(d\rho^0)^2-\mathcal{D}^2(d\rho_x^2+d\rho_y^2+d\rho_z^2 ).
\label{ds2-xyz}
\end{eqnarray}
It defines the metric tensor components $g_{ik}$ for the uniform coordinates $\rho^0$, $\rho^1$, $\rho^2$, $\rho^3$ and also $g^{ii}=1/g_{ii}$.

According to Eq.~$(\ref{def-N})$ $\mathcal{N}>1$ for $r_0^2>0$,  hence the dependence of $r$ on $\rho$ is monotonic. For asymptotically 
large $\rho \to 
\infty$, $\mathcal{D}\to 1$ in accordance with  Eq.~$(\ref{def-D})$, therefore  $r$ and $\rho$ are approximately equal to each other. 
The minimal value of $r$ equal to zero  corresponds to the minimal possible positive value of $\rho=\rho_{min}$. This value is the 
maximal number obeying the equation $\mathcal{D}(\rho)=0$  and is equal to
\begin{eqnarray} 
\rho_{min}=r_e/2-r_g/4. 
\label{rho-min}
\end{eqnarray}
This means that the sphere in the three-dimentional space
$(\rho_x,\rho_y,\rho_z)$ with the radius $\rho=\rho_{min}$ 
corresponds to the point $r=0$. There is no  contradiction in this respondency since according to Eq.~$(\ref{ds2-xyz})$ the 
distance between any two points on the sphere is zero due to the relation $\mathcal{D}(\rho_{min})=0$. 
\section{Tetrad representation}
\label{sect-4}
 In the tetrad representation proposed in Ref.~\cite{Moll-3}, the fundamental variables of the gravitational field are four unit four-vectors $h_{(a)}$ 
($a=0,\;1,\;2,\;3$ is a counting number of the four-vector)  being functions of the coordinates of  points in the four-dimensional 
spacetime. Their covariant and 
contravariant components are related  in the usual way by means of the metric tensor:
\begin{eqnarray}
h_{(a)i}=g_{ik}h^k_{(a)},\;\;\;h^i_{(a)}=g^{ik}h_{(a)k}.
\label{tetr-cc} 
\end{eqnarray}
The four-vector with $a=0$ is chosen time-like, while all others are space-like, namely
\begin{eqnarray}
h_{(a)i}h^i_{(b)}=\eta_{ab}
\label{tetr-scal}
\end{eqnarray}
with the diagonal matrix $\eta_{ab}=\rm{diag}(1,-1,-1,-1)$. Defining $\eta^{ab}$ equal to $\eta_{ab}$ the vectors $h^{(a)}$ 
can be expressed in terms of  
$h_{(a)}$ with the equation valid both for the covariant and contravariant components \cite{LL,Moll-3}
\begin{eqnarray}
h^{(a)i}=\eta^{ab}h^i_{(b)},\;\;h^{(a)}_i=\eta^{ab}h_{(b)i}.
\label{tetr-contr}
\end{eqnarray}
As follows from Eqs.~$(\ref{tetr-scal})$ and $(\ref{tetr-contr})$ the orthogonality conditions look like \cite{LL,Moll-3}
\begin{eqnarray}
h_{(a)i}h^{(b)i}=\delta_{a}^{b},
\label{orth-ab}\\
h_{(a)i}h^{(a)k}=\delta_{i}^{k},
\label{orth-ik}
\end{eqnarray}
where $\delta_{i}^{k}$ and $\delta_{a}^{b}$ denote the Kronecker symbols. The simplest consequence of Eq. $(\ref{orth-ik})$ is
the fundamental relation between the tetrads and the metric tensor \cite{LL,SW,Moll-3}
\begin{eqnarray}
h_{(a)i}h^{(a)}_l=g_{il},
\label{tetr-gil}\\
h^j_{(a)}h^{(a)k}=g^{jk}.
\label{tetr-gjk}
\end{eqnarray}

The partial derivative of $h^{(a)}_i$ over the uniform coordinate $\rho^j$ is denoted by $h^{(a)}_{i,j}$, 
while $h^{(a)}_{i;j}$ denotes the covariant derivative, where by definition,
\begin{eqnarray}
h^{(a)}_{i;k} = \frac{\partial h^{(a)}_{i}}{ \partial \rho^k}-\Gamma^m_{i k}h^{(a)}_{m} 
\equiv h^{(a)}_{i,k}-\Gamma^m_{i k}h^{(a)}_{m}
\label{cov-der}
\end{eqnarray}
with $\Gamma^m_{i k}$ being the Christoffel symbols. They are obtained with the help of standard formulas \cite{LL,SW}
\begin{eqnarray}
\Gamma^m_{i k}=g^{mn}\Gamma_{n,i k},
\label{Crist-up}\\
\Gamma_{n,i k}=\frac{1}{2}\Bigl [ \frac{\partial g_{nk}}{\partial \rho^i}+\frac{\partial g_{ni}}{\partial \rho^k}
-\frac{\partial g_{ik}}{\partial \rho^n} \Bigr ]. 
\label{Crist-down}
\end{eqnarray}  
Using Eq.~$(\ref{tetr-gil})$ for the metric tensor $g_{il}$ in terms of the tetrads and substituting Eqs.~$(\ref{Crist-up})$ and $(\ref{Crist-down})$
into Eq.~$(\ref{cov-der})$ one gets the formula of interest \cite{Moll-3}:
\begin{eqnarray}
\nonumber
h^{(a)}_{i;l}=\frac{1}{2}(h^{(a)}_{i,l}-h^{(a)}_{l,i})+
\\
+\frac{h^{(a)n}}{2}[h_{(b)l}(h^{(b)}_{i,n}-h^{(b)}_{n,i}) 
+h_{(b)i}(h^{(b)}_{l,n}-h^{(b)}_{n,l})].  
\label{cov-der-1}
\end{eqnarray} 

In the following, we need the formula for $h^{(a)k}_{\;\;\;\;\;;l}$. In order to obtain it, let us calculate the covariant derivative of 
Eq.~$(\ref{orth-ab})$ that is
\begin{eqnarray}
h_{(a)i;l}h^{(b)i}+ h_{(a)i}h^{(b)i}_{\;\;\;\;\;;l}=0.
\label{cov-delab}
\end{eqnarray}   
Multiplying this equation by $h^{(a)k}$, summing over $a$, and making use of Eq.~$(\ref{orth-ik})$ we get
\begin{eqnarray}
h^{(b)k}_{\;\;\;\;\;;l}=-h^{(b)i}h^{(a)k} h_{(a)i;l}
=-h^{(b)i}h_{(a)}^k h^{(a)}_{i;l}.
\label{cov-hkl}
\end{eqnarray}   
 Substituting Eq.~$(\ref{cov-der-1})$ into Eq.~$(\ref{cov-hkl})$, summing over $a$, and taking into account 
Eq.~$(\ref{tetr-gjk})$ we get finally
\begin{eqnarray}
\nonumber
h^{(b)k}_{\;\;\;\;\;;l}=\frac{1}{2}h^{(b)m}h_{(a)}^k (h^{(a)}_{l,m}-h^{(a)}_{m,l})
+\frac{1}{2}g^{ik}(h^{(b)}_{i,l}-h^{(b)}_{l,i})+
\\
+\frac{1}{2}h^{(b)m}g^{ik}h_{(a)l}(h^{(a)}_{i,m}-h^{(a)}_{m,i}).
\label{cov-hkl-fin}   
\end{eqnarray}
As seen from Eqs.~$(\ref{cov-der-1})$ and $(\ref{cov-hkl-fin})$ the covariant derivatives $h^{(a)}_{i;l}$ and $h^{(b)k}_{\;\;\;\;\;;l}$ depend 
on the partial derivatives $h^{(a)}_{m,i}$ linearly.
\section{Lagrangian for the Reissner--\\Nordstr\"om solution}
\label{sect-5}
     The formula for the total lagrangian density reads
\begin{eqnarray}
\mathcal{L}_{tot}=\mathcal{L}_{g}+\mathcal{L}_{em},
\label{tot-lagr}
\end{eqnarray}
where $\mathcal{L}_{g}$ is the lagrangian density of gravitational field, 
while $\mathcal{L}_{em}$ denotes the lagrangian density of electromagnetic field. In the tetrad representation, the former is given by the formula 
\cite{Moll-3,Moll-4}
\begin{eqnarray}
\mathcal{L}_{g}=\frac{|h|}{2 \kappa}\Bigl ( h^k_{(a);l}h^{(a)l}_{\;\;\;\;\;;k}
-h^k_{(a);k}h^{(a)l}_{\;\;\;\;\;;l} \Bigr ), 
\label{gra-lagr}                              
\end{eqnarray}  
where 
\begin{eqnarray}
\kappa=8\pi k/c^4, 
\label{def-kappa}
\end{eqnarray}
and $|h|$ denotes the determinant of the $4 \times 4$ matrix $h_{(a)i}$. 
As is demonstrated by  Eq.~$(\ref{cov-der-1})$ or $(\ref{cov-hkl-fin})$ the covariant derivatives of the tetrad components depend
 only on  the tetrad components  and their first partial derivatives with respect to the coordinates $\rho^m$. Therefore the lagrangian $\mathcal{L}_{g}$ 
in Eq.~$(\ref{gra-lagr})$ depends also on the tetrads and their 
first  partial derivatives.

According to Eq.~$(\ref{ds2-xyz})$ and relation \cite{Moll-3,Moll-4}
\begin{eqnarray}
g=-|h|^2,
\label{det-g-h}
\end{eqnarray}
the formula for
$|h|$ for the uniform coordinates for the RN solution is
\begin{eqnarray}
|h|=\mathcal{N}\mathcal{D}^2.
\label{det-h-nd}
\end{eqnarray}  

The general formula for the lagrangian density of the electromagnetic field reads \cite{LL,SW}
\begin{eqnarray}
\mathcal{L}_{em}=-\frac{|h|}{16 \pi}F_{ik}F^{ik}.
\label{lem} 
\end{eqnarray}
Since $F_{ik}F^{ik}$ is invariant it is possible to use Eq.~$(\ref{sol-F01})$ for the electromagnetic tensor expressing
 $r$ with the help of  Eq.~$(\ref{r-rho})$. Using also Eq.~$(\ref{det-h-nd})$ the final formula for $\mathcal{L}_{em}$ 
for the RN solution in the uniform coordinates is obtained
\begin{eqnarray}
\mathcal{L}_{em}=\frac{e^2\mathcal{N}}{8 \pi \rho^4\mathcal{D}^2}=\frac{\mathcal{N}r_e^2}{\kappa \rho^4\mathcal{D}^2}.
\label{lem-uni}
\end{eqnarray}  
In the second representation for $\mathcal{L}_{em}$ in  Eq.~$(\ref{lem-uni})$ more convenient below, the formula
\begin{eqnarray}
\frac{e^2}{8 \pi}=\frac{r^2_e}{\kappa}
\label{e-re-kap} 
\end{eqnarray}
is taken into account that follows from definitions given by Eqs.~$(\ref{re})$ and $(\ref{def-kappa})$.

For the tetrads  with the covariant components
\begin{eqnarray}
h_{(0)k}= h_k^{(0)}=\frac{\mathcal{N}}{\mathcal{D}}\delta_k^0,\;\;\;h_{(\mu)k} =-h_k^{(\mu)}=\mathcal{D}\delta_k^{\mu}
\label{rn-tetr} 
\end{eqnarray}
for any $k$ and $\mu$, relation~$(\ref{tetr-gil})$ is fulfilled. Formulas for the contravariant components 
follow from Eqs.~$(\ref{tetr-cc})$,  $(\ref{rn-tetr})$, and also from
Eq.~$(\ref{ds2-xyz})$ for the metric tensor
\begin{eqnarray}
h^{(0)i}= h_{(0)}^i=g^{00}h_{(0)0}\delta^i_0=\delta^i_0 \mathcal{D}/\mathcal{N},
\label{h-up-0-down-0}\\
h_{(\mu)}^{i}=-h^{(\mu)i}
=g^{i k}h_{(\mu)k}
=-\delta^{i}_{\mu}/\mathcal{D}.
\label{hlm-obv}
\end{eqnarray}
These contravariant components of the tetrad obey Eq.~$(\ref{tetr-gjk})$.
Using formulas for $h_{\;\;\;\;\;:l}^{(a)k}$ obtained in Subsec. \ref{appendix-2}
 of Appendix one gets
\begin{eqnarray}
\mathcal{L}_g=\frac{\mathcal{N}\mathcal{D}'}{\kappa\mathcal{D}}\Bigl \{ 2 \frac{\mathcal{N}'}{\mathcal{N}}-\frac{\mathcal{D}'}{\mathcal{D}}
\Bigr \},
\label{lg-nd}  
\end{eqnarray}
where $\mathcal{N}'$ and $\mathcal{D}'$ denote derivatives of $\mathcal{N}(\rho)$ and $\mathcal{D}(\rho)$ with respect to $\rho$.

Substituting $\mathcal{L}_{em}$ given by Eq.~$(\ref{lem-uni})$ and $\mathcal{L}_g$ from Eq.~$(\ref{lg-nd})$ into Eq.~$(\ref{tot-lagr})$ and using 
Eqs.~$(\ref{def-D})$ and  $(\ref{def-N})$ respectively for $\mathcal{D}(\rho)$ and 
$\mathcal{N}(\rho)$ we obtain for the total lagrangian 
density the very simple formula
\begin{eqnarray}
\mathcal{L}_{tot}=\frac{r^2_0}{\kappa \rho^4}.
\label{ltot-nd}   
\end{eqnarray}
Since the three-dimentional space $(\rho_x,\rho_y,\rho_z)$ consists of points for which
$\rho=\sqrt{\rho_x^2+\rho_y^2+\rho_z^2}$ obeys the inequality $\rho\geq \rho_{min}$
the value of the total lagrangian is given by the equation
\begin{eqnarray}
L_{tot}=\int_{\rho\geq \rho_{min}}\mathcal{L}_{tot}d\rho_x d\rho_y d\rho_z=\frac{4 \pi  r^2_0}{\kappa \rho_{min}}.
\label{ltot-int}
\end{eqnarray}  
Making use of Eqs.~(\ref{rg}--\ref{r0})  and $(\ref{rho-min})$ we get the final result:
\begin{eqnarray}
L_{tot}=c^2\sqrt{\frac{e^2}{k}}+mc^2.
\label{ltot-final}
\end{eqnarray}
This formula shows that the lagrangian in the tetrad representation and the action $S=L_{tot}t$ are finite in spite of the singular behaviour of the 
electromagnetic
and gravitational fields near the point $r=0$ ($\rho=\rho_{min}$). 

It is not the case when we consider the metric tensor components
as the fundamental variables describing the gravitational field. Indeed, it is well known \cite{LL,SW,Hilb,AE-1} that the scalar curvature
 $\mathcal{R}$ and hence the lagrangian density of the gravitational field 
\begin{eqnarray}
\tilde{\mathcal{L}}_{g}=-\frac{\mathcal{R}\sqrt{-g}}{2\kappa}
\label{lgra-curv}
\end{eqnarray}
is zero if the matter is represented by the electromagnetic field only. Since the 
lagrangian density for the electromagnetic field 
defined by Eqs.~$(\ref{lem})$ and $(\ref{sol-F01})$ has nonintegrable singularity at $r=0$ the total lagrangian 
corresponding to Eqs.~(\ref{tot-lagr}, \ref{lem}), and 
$(\ref{lgra-curv})$ is meaningless, while Eq.~$(\ref{ltot-final})$ provides the finite lagrangian in the tetrad 
representation. We assume
 that it is these sixteen functions $h^{(a)}_i(\rho^k)$ of the coordinates $\rho^k$ which are the fundamental gravitational variables rather than 
the components of the metric tensor. The knowledge of a true lagrangian density is of crucial importance since it could be used in calculations
of the process amplitudes in quantum theory with the help of the Feynman functional integrals \cite{RPF-1,FP}.
\section{Equations of motion}
\label{sect-6}
     Remarkable formula $(\ref{ltot-final})$ follows from the choice of the tetrad defined by Eqs.~(\ref{rn-tetr}--\ref{hlm-obv}) 
that reproduces the metric tensor corresponding to the RN
 solution. But the true tetrads are to obey the Lagrange equations
\begin{eqnarray}
\frac{\partial \mathcal{L}_{tot}}{\partial h^{(c)}_p}=\frac{\partial}{\partial \rho^{q}} \Bigl [
\frac{\partial \mathcal{L}_{tot}}{\partial h^{(c)}_{p,q}}\Bigr ].
\label{lagr-eq} 
\end{eqnarray}
The total lagrangian density depends on the variables $h^{(c)}_p$ describing the gravitational field and their partial 
derivatives $h^{(c)}_{p,q}$ with respect to $\rho^q$. The variables for the electromagnetic field are the four-potentials $A_{i}$. The 
electromagnetic field tensor $F_{ik}$ 
\begin{eqnarray}
F_{ik}=A_{k,i}-A_{i,k}\equiv \frac{\partial A_k}{\partial \rho^i}-\frac{\partial A_i}{\partial \rho^k}
\label{Fik-aik}
\end{eqnarray}
is a linear combination of the partial derivatives of $A_{i}$ with respect to the spacetime coordinates.
Formula $(\ref{lem})$ for the lagrangian density of the electromagnetic field rewritten in the form
\begin{eqnarray}
\mathcal{L}_{em}=-\frac{|h|}{16 \pi}g^{ij}g^{kl}F_{ik}F_{jl}
\label{l-em-gFij}
\end{eqnarray}
shows that $\mathcal{L}_{em}$
depends on $A_{i,k}$  and $h^{(c)}_p$ since the determinant $|h|$ and the metric tensor can be expressed in terms of
the tetrad components according to Eq.~$(\ref{tetr-gjk})$, therefore $\mathcal{L}_{em}$ does not depend on $h^{(c)}_{p,q}$. 
The only lagrangian density, which depends on the
derivatives $h^{(c)}_{p,q}$, is that of the gravitational field $\mathcal{L}_{g}$ defined by Eq.~$(\ref{gra-lagr})$.

Since $\mathcal{L}_{g}$ 
is a sum of bilinear products of $h^{(a)k}_{\;\;\;\;\;;l}$ we need the 
derivative of $h^{(a)k}_{\;\;\;\;\;;l}$ with respect to $h^{(c)}_{p,q}$. Making use of Eq.~$(\ref{cov-hkl-fin})$ it is easy to get 
the result
\begin{eqnarray}
\nonumber
\frac{\partial h^{(b)k}_{\;\;\;;l}}{\partial h^{(c)}_{p,q}}=\frac{1}{2}h^k_{(c)}(\delta^p_lh^{(b)q}-\delta^q_l h^{(b)p})+ \\
+\frac{1}{2}h_{(c)l}(g^{pk}h^{(b)q}- g^{qk}h^{(b)p})+\frac{1}{2}\delta^b_c(g^{pk}\delta^q_l-g^{qk}\delta^p_l).
\label{dhbkl-dhcpq} 
\end{eqnarray}
Multiplying $h^{l}_{(a);k}$ from Eq.~$(\ref{cov-hkl-fin})$ by $\partial h^{(a)k}_{\;\;\;\;\;;l}/\partial h^{(c)}_{p,q}$ from 
Eq.~$(\ref{dhbkl-dhcpq})$, after some algebra,  one gets the simple expression 
\begin{eqnarray}
\frac{\partial \{h^{(a)k}_{\;\;\;\;\;;l} h^{l}_{(a);k}\}}{\partial h^{(c)}_{p,q}}=2 h_{(c)l}\gamma^{qpl}
\label{lag1-dhcpq}
\end{eqnarray}
in terms of the tensor $\gamma^{qpl}$, where its covariant components are defined by \cite{Moll-3,Moll-4}
\begin{eqnarray}
\gamma_{ikl}= h_{(a)i}h^{(a)}_{k;l}=-h_{(a)k}h^{(a)}_{i;l}=-\gamma_{kil}.
\label{g_ikl}
\end{eqnarray}

Taking $k=l$ in Eq.~$(\ref{dhbkl-dhcpq})$ and summing over $k$ we get
\begin{eqnarray}
\frac{\partial h^{(b)k}_{\;\;\;\;\;;k}}{\partial h^{(c)}_{p,q}}=h^p_{(c)}h^{(b)q}-h^q_{(c)}h^{(b)p}.
\label{dhbkk-dhcpq}
\end{eqnarray}
Let us define the four-vector $\Phi^q$ by the relation
\begin{eqnarray}
\Phi^q=\gamma^{lq}_{\;\;\;l}.
\label{def-phiq}
\end{eqnarray}
Then, taking $k=l$ in Eq.~$(\ref{cov-hkl-fin})$, presenting $g^{ik}$ as the bilinear product of the tetrads with the help of Eq.~$(\ref{tetr-gjk})$,
and summing over $l$, we get \cite{Moll-3} 
\begin{eqnarray}
h^{(b)l}_{\;\;\;\;\;;l}=-h^{(b)m}\Phi_m.
\label{hbrr}
\end{eqnarray}
The simplest consequence of Eqs.~$(\ref{dhbkk-dhcpq})$ and $(\ref{hbrr})$ is the formula
\begin{eqnarray}
\frac{\partial \{h^{(b)k}_{\;\;\;\;\;;k} h^{l}_{(b);l}\}}{\partial h^{(c)}_{p,q}}=2(h^q_{(c)}\Phi^{p}-h^p_{(c)}\Phi^{q}).
\label{lg2-dhcpq}
\end{eqnarray}
Substituting Eqs.~$(\ref{lag1-dhcpq})$ and $(\ref{lg2-dhcpq})$ into Eq.~$(\ref{gra-lagr})$ we get finally
\begin{eqnarray}
\frac{\partial}{\partial \rho^{q}} \Bigl [
\frac{\partial \mathcal{L}_{tot}}{\partial h^{(c)}_{p,q}}\Bigr ]=
\frac{\partial}{\partial \rho^{q}} \Bigl [\frac{|h|}{\kappa}\Bigl (h^l_{(c)}\gamma^{qp}_{\;\;\;l} 
+h^p_{(c)} \Phi^q-h^q_{(c)} \Phi^p \Bigr) \Bigr].
\label{righ-lagr}
\end{eqnarray}    

In order to obtain the left-hand side of Eq.~$(\ref{lagr-eq})$, one needs the formula  for 
$\partial \mathcal{L}_{tot}/\partial h^{(c)}_{p}$, 
where
\begin{eqnarray}
\frac{\partial \mathcal{L}_{tot}}{\partial h^{(c)}_{p}}=\frac{\partial \mathcal{L}_{g}}{\partial h^{(c)}_{p}}
+\frac{\partial \mathcal{L}_{em}}{\partial h^{(c)}_{p}}.
\label{lgtot-dhcp}
\end{eqnarray} 
Using Eq.~$(\ref{gra-lagr})$ we get
\begin{eqnarray}
\nonumber
\frac{\partial \mathcal{L}_{g}}{\partial h^{(c)}_{p}}=\frac{|h|}{\kappa}
\Bigl \{ h^{(b)l}_{\;\;\;\;\;;k}\frac{\partial h^{(b)k}_{\;\;\;\;\;\;;l}}{\partial h^{(c)}_{p}}
-h^{(b)l}_{\;\;\;\;\;;l}\frac{\partial h^{(b)k}_{\;\;\;\;\;\;;k}}{\partial h^{(c)}_{p}}\Bigr \}+ \\
+\frac{1}{2\kappa}
\Bigl \{ h^k_{(a);l}h^{(a)l}_{\;\;\;\;\;;k}-h^k_{(a);k}h^{(a)l}_{\;\;\;\;\;;l}\Bigr \} 
\frac{\partial |h|}{\partial h^{(c)}_{p}}.
\label{lggra-dhcp}
\end{eqnarray}
Equation $(\ref{lggra-dhcp})$ contains
 $\partial h^{(b)k}_{\;\;\;;l}/\partial  h^{(c)}_{p}$ 
that can be obtained 
from Eq.~$(\ref{cov-hkl-fin})$. Differentiating  for this aim Eq.~$(\ref{orth-ab})$ with respect to $h^{(c)}_{p}$ and using the  trivial formulas
\begin{eqnarray}
\frac{\partial h^{(b)}_{m}}{\partial h^{(c)}_{p}}=\delta^b_c\delta^p_m,\;\;\frac{\partial h_{(b)m}}{\partial h^{(c)}_{p}}=\eta_{bc}\delta^p_m
\label{triv-form}
\end{eqnarray}  
we get 
\begin{eqnarray}
\frac{\partial h^j_{(b)}}{\partial h^{(c)}_{p}}=-h^p_{(b)} h^j_{(c)},
\label{der-1}\\
\frac{\partial h^{(b)j}}{\partial h^{(c)}_{p}}=-h^{(b)p}h^j_{(c)}.
\label{der-2}
\end{eqnarray}
Differentiating the left-hand side of Eq.~$(\ref{tetr-gjk})$ and taking into account Eqs.~$(\ref{der-1},\ref{der-2})$ we get the formula for the  derivative  
of $g^{jl}$ with respect to $h^{(a)}_{m}$
\begin{eqnarray}  
\frac{\partial g^{jl}}{\partial h^{(a)}_{m}}=-g^{jm}h^l_{(a)} -g^{ml}h^j_{(a)}.
\label{der-gik}
\end{eqnarray}
Making use of Eqs.~(\ref{triv-form}--\ref{der-gik})  
in the calculation of the partial derivative of $h^{(b)k}_{\;\;\;;l}$ over 
$h^{(c)}_{p}$, we get after some algebra
\begin{eqnarray}
\nonumber
\frac{\partial h^{(b)k}_{\;\;\;\;;l}}{\partial h^{(c)}_{p}}= \frac{1}{2}h^{(b)p}h^i_{(c)}(\gamma^k_{\;\;il}
+\gamma_{li}^{\;\;\;k})
+h^k_{(c)}h^{(b)i}\gamma^p_{\;\;il} - \\ 
-\frac{1}{2}g^{pk}h^i_{(c)}h^{(b)m}(\gamma_{lim}+\gamma_{mil})
-\frac{1}{2} \delta^p_l h^{(b)m}h^i_{(c)}(\gamma^k_{\;im}+\gamma_{im}^{\;\;\;k}).
\label{der-hbkl-hcp}
\end{eqnarray}

In order to express the derivative 
$\partial h^{(b)k}_{\;\;\;;l}/\partial h^{(c)}_{p}$ 
in terms of the tensor $\gamma_{kil}$, 
its definition by Eq.~$(\ref{g_ikl})$ is used and the formula 
\begin{eqnarray}
h^{(b)}_{i,l}-h^{(b)}_{l,i} = h^{(b)}_{i;l}-h^{(b)}_{l;i}=h^{(b)m}(\gamma_{mil}-\gamma_{mli}) 
\label{hil-hli}
\end{eqnarray}   
is applied. Formula $(\ref{hil-hli})$ follows from the obvious transformation using  Eq.~$(\ref{orth-ab})$ and the definition of $\gamma_{ikl}$ by 
 Eq.~$(\ref{g_ikl})$
\begin{eqnarray}
h^{(b)}_{i;l}=\delta^b_a h^{(a)}_{i;l}=h^{(b)m}h_{(a)m}h^{(a)}_{i;l}=h^{(b)m}\gamma_{mil}.
\label{tr-hil-hli}
\end{eqnarray}
Making use of Eq.~$(\ref{hil-hli})$ the basic formula $(\ref{cov-hkl-fin})$ can easily be rewritten as \cite{Moll-3}
\begin{eqnarray}
h^{(b)k}_{\;\;\;\;\;;l}=-h^{(b)m}\gamma^k_{\;ml}.
\label{hkl-gamma}
\end{eqnarray}
Combining Eq.~$(\ref{hkl-gamma})$ with Eq.~$(\ref{der-hbkl-hcp})$ one gets
\begin{eqnarray}
h^{l}_{(b)\;;k}\frac{\partial h^{(b)k}_{\;\;\;\;\;\;;l}}{\partial h^{(c)}_{p}}=h^i_{(c)}\gamma^{pml}(\gamma_{mli}+\gamma_{lim}). 
\label{hlk-dhkl-gamma}
\end{eqnarray}

Taking in  Eq.~$(\ref{der-hbkl-hcp})$ $k=l$ and summing over $k$ we get
\begin{eqnarray}
\frac{\partial h^{(b)k}_{\;\;\;\;\;;k}}{\partial h^{(c)}_{p}}=h^{(b)p}h^i_{(c)}\Phi_i+h^i_{(c)}h^{(b)m}(\gamma^p_{\;mi}-\gamma^p_{\;im}). 
\label{hbkk-dhcp-gamma}
\end{eqnarray}
Combining this formula with  Eq.~$(\ref{hbrr})$ we get
\begin{eqnarray}
h^{(b)s}_{\;\;\;\;\;;s}
\frac{\partial h^{(b)k}_{\;\;\;\;\;;k}}{\partial h^{(c)}_{p}}=h_{(c)i}\Bigl [ (\gamma^{pim}-\gamma^{pmi})\Phi_m -\Phi^p\Phi^i \Bigr ].
\label{hlk-dhrr-gamma}
\end{eqnarray}

The last formula needed to obtain 
$\partial \mathcal{L}_{g}/\partial h^{(c)}_{p}$ 
is the following \cite{Moll-3,Moll-5}:
\begin{eqnarray}
\frac{\partial  |h|}{\partial h^{(c)}_{p}}=|h|h^p_{(c)}.
\label{der-determ}
\end{eqnarray}
Finally, substituting $h^{l}_{(b)\;;k}\partial h^{(b)k}_{\;\;\;\;\;\;;l}/\partial h^{(c)}_{p}$ given by
Eq.~$(\ref{hlk-dhkl-gamma})$, 
\linebreak
$h^{(b)s}_{\;\;\;\;\;;s}\partial h^{(b)k}_{\;\;\;\;\;;k}/\partial h^{(c)}_{p}$
from Eq.~$(\ref{hlk-dhrr-gamma})$, and $\partial  |h|/\partial h^{(c)}_{p}$ given by Eq.~$(\ref{der-determ})$ 
into  Eq.~$(\ref{lggra-dhcp})$ we obtain 
the formula 
\begin{eqnarray}
\nonumber
\frac{\partial \mathcal{L}_{g}}{\partial h^{(c)}_{p}}=\frac{|h|}{\kappa}\Bigl \{ h^i_{(c)}\Bigl [ \gamma^{pml}
\Bigl (\gamma_{mli}+\gamma_{lim}\Bigr )+ \Phi_i\Phi^p +
 \\
+ \Bigl (\gamma_i^{\;\;pm}+\gamma^{pm}_{\;\;\;\;i} \Bigr ) \Phi_m\Bigr ] 
+\frac{h^p_{(c)}}{2}\Bigl [ \gamma^{kml}\gamma_{lmk}-\Phi_m \Phi^m\Bigr ]\Bigr \}.
\label{fin-lg-der-hap}
\end{eqnarray}

In the calculation of $\partial \mathcal{L}_{em}/\partial h^{(c)}_{p}$, where 
\begin{eqnarray}
\frac{\partial \mathcal{L}_{em}}{\partial h^{(c)}_{p}}= -\frac{F_{ik}F_{jl}}{16 \pi}\Bigl \{ g^{ij}g^{kl}
\frac{\partial  |h|}{\partial h^{(c)}_{p}}+|h| \frac{\partial (g^{ij}g^{kl})}{\partial h^{(c)}_{p}} \Bigr \},
\label{lem-der-hap} 
\end{eqnarray}
Eq.~$(\ref{l-em-gFij})$ for $\mathcal{L}_{em}$ is used and
the independence of the covariant components of the electromagnetic field tensor $F_{ik}$ on $h^{(c)}_{p}$ is taken into account.
Substituting $\partial g^{jl}/\partial h^{(c)}_p$
from Eq.~$(\ref{der-gik})$ and $\partial |h|/\partial h^{(c)}_p$ from Eq.~$(\ref{der-determ})$ into 
Eq.~$(\ref{lem-der-hap})$ we get
\begin{eqnarray}
\frac{\partial \mathcal{L}_{em}}{\partial h^{(c)}_{p}}=-|h|T^p_i h^i_{(c)},
\label{lem-var-hap}
\end{eqnarray}
where $T^p_i$ is the energy-momentum tensor of the electromagnetic field defined by Eq.~$(\ref{Tik})$.

Substituting  $\partial \mathcal{L}_{g}/\partial h^{(c)}_{p}$ and $\partial \mathcal{L}_{em}/\partial h^{(c)}_{p}$ respectively from Eqs.~$(\ref{fin-lg-der-hap})$ and 
$(\ref{lem-var-hap})$ into Eq.~$(\ref{lgtot-dhcp})$ and taking into account Eq.~$(\ref{Tik})$ 
for $T^p_i$ we get
for the left-hand side of motion equation $(\ref{lagr-eq})$ the final formula
\begin{eqnarray}
\nonumber
\frac{\partial \mathcal{L}_{tot}}{\partial h^{(c)}_{p}}=\frac{|h|}{4 \pi}F^{pl}F_{il}h^i_{(c)}+h^p_{(c)}\mathcal{L}_{tot} \\ 
+\frac{|h|}{\kappa}h^i_{(c)} \Bigl \{ \gamma^{pml}\Bigl ( \gamma_{mli}+\gamma_{lim}\Bigr ) 
+\Bigl (\gamma^{pm}_{\;\;\;\;i}+\gamma_i^{\;\;pm}  \Bigr ) \Phi_m +\Phi_i \Phi^p \Bigr \}.
\label{fin-ltot-var-hap}   
\end{eqnarray}
Here, $\mathcal{L}_{tot}$ is described by very short formula $(\ref{ltot-nd})$ which simplifies any calculations.
Note that expression~$(\ref{Tik})$ for $T^p_i$ contains the term proportional to the product of the Kronecker symbol and the lagrangian density 
$\mathcal{L}_{em}$ 
given by Eq.~$(\ref{lem})$. The lagrangian density $\mathcal{L}_{g}$ for the gravitational field can be rewritten in the form \cite{Moll-3,Moll-4}
\begin{eqnarray}
\mathcal{L}_{g}=\frac{|h|}{2 \kappa}[\gamma^{kml}\gamma_{lmk}-\Phi_m \Phi^m].
\label{lg-gamma-phi}
\end{eqnarray}
The sum of the two last terms in the right-hand side of Eq.~$(\ref{fin-lg-der-hap})$
proportional to $\mathcal{L}_{g}$  and the term in Eq.~$(\ref{lem-var-hap})$
 proportional to $\mathcal{L}_{em}$  
gives the term proportional to $\mathcal{L}_{tot}$ in Eq.~$(\ref{fin-ltot-var-hap})$.  

Thus, the equation of motion  for $h^{(c)}_p$ is Lagrange 
equation~$(\ref{lagr-eq})$, where its left-hand side is given by Eq.~$(\ref{fin-ltot-var-hap})$, while the 
right-hand side is given by Eq.~$(\ref{righ-lagr})$. It is shown in Subsec. \ref{appendix-4} of Appendix, 
that the tetrad $h^{(c)}_p$ defined by Eqs.~$(\ref{rn-tetr})$ obeys the Lagrange equation. 
\section {Conservation of energy-momentum \\four-vector}
\label{sect-7}
In the tetrad formalism, the total energy-momentum pseudo-tensor density, $\mathcal{T}_{i}^{k}$ is related to the superpotential 
$U_{i}^{\;\;kl}$, proposed by Moller (see Refs.~\cite{Moll-3,Moll-4,Moll-5}), where 
\begin{eqnarray}
U_{i}^{\;\;kl}=-U_{i}^{\;\;lk}=h^{(a)k}\frac{\partial \mathcal{L}_{tot}}{\partial h^{(a)i}_{\;\;\;\;,l}}
=h^{(a)}_{i} \frac{\partial \mathcal{L}_{tot}}{\partial h^{(a)}_{l,k}},
\label{def-U} 
\end{eqnarray}
with the formula
\begin{eqnarray}
\mathcal{T}_{i}^{k}=U_{i\;\;\;\; ,l}^{\;\;kl} \equiv \frac{\partial U_{i\;\;\;\; }^{\;\;kl}}{\partial \rho^l}.
\label{Tik-U}
\end{eqnarray}
It is shown in Refs.~\cite{Moll-3,Moll-4,Moll-5} that
\begin{eqnarray}
U_{i}^{\;\;k l}=\frac{|h|}{\kappa}\Bigl \{ h_{(a)}^{k}  h^{(a)l}_{\;\;\;\;\; ;i}
+\Bigl ( \delta ^{k}_{i}h^{(a)l}-\delta^{l}_{i}h^{(a)k} \Bigr )h_{(a);s}^{s}\Bigr \},
\label{Moll-U}
\end{eqnarray}
or in terms of the tensor $\gamma_{ikl}$ and the four-vector $\Phi^{l}$ defined respectively by Eq.~$(\ref{g_ikl})$ and $(\ref{def-phiq})$
\begin{eqnarray}
U_{i}^{\;\;k l}=\frac{|h|}{\kappa}\Bigl \{ \gamma^{kl}_{\;\;\;i}
-\delta ^{k}_{i}\Phi^{l}+\delta^{l}_{i}\Phi^{k} \Bigr \}. 
\label{Moll-U1}
\end{eqnarray}

As seen from Eq.~$(\ref{Moll-U})$ Moller's superpotential is the tensor density 
under arbitrary coordinate transformations since it is  expressed
in terms of the tetrad vectors $h^{(a)k}$, their covariant derivatives $h^{(a)l}_{\;\;\;\;\; ;i}$, and the determinant $|h|$. 
The superpotential $U_{i}^{\;\;k l}$ is antisymmetric with respect to the indexes $k$ and $l$, therefore the divergence 
of the total energy-momentum pseudo-tensor density is zero
\begin{eqnarray}
\mathcal{T}_{i\; ,k}^{k}=U_{i\;\;\;\; ,l,k}^{\;\;kl}\equiv 
\frac{\partial ^2U_{i}^{\;\;k l}}{\partial \rho^{k} \partial \rho^{l}}=0.
\label{div-Tik}
\end{eqnarray}
As is well known \cite{LL,SW,AE-1,Moll-3,Moll-4,Moll-5} the conservation of the energy-momentum four-vector is a consequence of Eq.~$(\ref{div-Tik})$.

If the total energy-momentum pseudo-tensor 
is localized in the compact three-dimen\-si\-onal region  
$\mathcal{V}$ 
 such a system will be called  the insular one.
Then the metric tensor $g_{ik}$ for the insular system goes to its Minkowski limit $\eta_{ik}$ at $\rho \to \infty$ as
\begin{eqnarray}
g_{ik}(\rho)=\eta_{ik}(\rho)+O_1(\rho).
 \label{gik-etaik}
\end{eqnarray} 
Here, $O_n$ denotes the quantity which main term at $\rho \to \infty$ is proportional to $\rho^{-n}$ for $n=1,\;2,...\;$.
The standard consideration shows that the energy-momentum four-vector $P_i$ defined by
\begin{eqnarray}
P_i=\frac{1}{c}\int_{\mathcal{V}} \mathcal{T}_{i}^{0}d^3\rho
\label{Pi-Tik}
\end{eqnarray}
is conserved if all fields are zero outside the region $\mathcal{V}$, where $d^3\rho=d\rho^1d\rho^2d\rho^3\equiv d\rho_xd\rho_yd\rho_z$ is 
the volume 
element. Since the formula for $P_i$ can be rewritten in terms of the 
superpotential components according to Eq.~$(\ref{Tik-U})$
it can be expressed as the surface integral using the Gauss theorem
\begin{eqnarray}
P_i=\frac{1}{c}\int_{\mathcal{V}} \frac{\partial U_{i}^{\;\;0 \lambda}}{\partial \rho^{\lambda}}d^3\rho
=\frac{1}{c}\int_{\Sigma}U_{i}^{\;\;0 \lambda}k_{\lambda}d \sigma.
\label{Pi-Uikl}
\end{eqnarray}
Here, $\Sigma$ is the closed surface enclosing the region  $\mathcal{V}$, while the three-dimensional vector $k_{\lambda}$ is the unit  outer normal to the surface. 
The element of the surface is defined by
\begin{eqnarray}
k_{\lambda}d \sigma=\epsilon_{\lambda \mu \nu}d \rho^{\mu}\delta \rho^{\nu},
\label{dsigma}
\end{eqnarray}
where $\epsilon_{\lambda \mu \nu}$ is the totally antisymmetric three-dimensional Levi-Civita  symbol, while $d \rho^{\mu}$ and 
$\delta \rho^{\nu}$ are infinitesimal three-vectors on the surface $\Sigma$.

When the matter fields are localized mainly in the region $\mathcal{V}$ and $\mathcal{T}_{i}^{k}$ goes to zero outside $\mathcal{V}$ at $\rho \to \infty$ as  
a quantity 
of $O_n$ with $n \geq 4$ the system will be also called the insular system.
For this case, the full three-dimensional space ($\mathcal{V}_{\infty}$) 
is to be considered and for the surface $\Sigma$, we will choose $\Omega_{\mathcal{R}}$ with $\mathcal{R} \to \infty$. Here $\Omega_{\mathcal{R}}$ denotes the surface 
of the sphere with the radius $\mathcal{R}$.
The surface integral in Eq.~$(\ref{Pi-Uikl})$ becomes the limit of the 
integral on $\Omega_{\mathcal{R}}$ at $\mathcal{R} \to \infty$. 

Substituting
formulas $(\ref{det-h-nd})$ for $|h|$, $(\ref{h-up-0-down-0})$ and 
$(\ref{hlm-obv})$ for 
$h_{(a)}^{k}$, and (\ref{h0-0m}--\ref{hn-mn})
for $h^{(a)l}_{\;\;\;\;\; ;i}$, obtained in Appendix, into  Eq.~$(\ref{Moll-U})$ 
we get for the nonzero components of the superpotential $U_{i}^{\;\;k l}$ for the RN
 solution 
\begin{eqnarray}
U_{0}^{\;\;0 \lambda}=-U_{0}^{\;\; \lambda 0}=-2\mathcal{N} \frac{n_{\lambda} \mathcal{D}'}
{\kappa\mathcal{D}},
\label{W0-0lam} 
\end{eqnarray}
where $n_{\lambda}$ is the unit three-vector
\begin{eqnarray}
n_{\lambda}=\rho^{\lambda}/\rho
\label{n_lam} 
\end{eqnarray}
with $\rho^{\lambda}$ defined by Eqs.~(\ref{def-rhoxyz}).
For $\mu \neq \lambda$, one gets
\begin{eqnarray}
U_{\mu}^{\;\;\lambda \mu}=-U_{\mu}^{\;\;\mu \lambda}=
\frac{n_{\lambda}}{\kappa} \mathcal{N}'.
\label{Wmu-lamu}
\end{eqnarray}
Note that for the RN
 solution
the full three-dimensional space for the uniform coordinates $\rho_x,\;\rho_y,\;\rho_z$ consists of all points with 
$\rho=\sqrt{\rho^2_x+\rho^2_y+\rho^2_z}\geq \rho_{min}$, where $\rho_{min}$ is 
given by Eq.~$(\ref{rho-min})$ and $\mathcal{D}(\rho_{min})=0$. This means that the surface integral consists 
of the integrals over the spheres with the radii $\rho=\rho_{min}$ and $\rho=\mathcal{R}$  with $\mathcal{R} \to \infty$. As seen 
from Eq.~$(\ref{W0-0lam})$  all the superpotential
components $U_0^{\;\;0 \lambda}$ are infinite  on the sphere with the radius $\rho_{min}$ since $\mathcal{D}(\rho_{min})=0$
and $U_0^{\;\;0 \lambda}$  behave near $\rho_{min}$ as   $1/(\rho-\rho_{min})$.
This makes the definition of the energy given by Eq.~$(\ref{Pi-Uikl})$ meaningless for any parameter $m$ obeying inequality in 
Eq.~$(\ref{cond-1})$ since the energy is infinitely large. Also, there is no solution with $e$ and $m$ obeying Eq.~$(\ref{cond-1})$
with a finite total inertial mass which can be used as a model for electrons.      

But it is not the case when the limit $m \to\sqrt{e^2/k}$ is considered, and
\begin{eqnarray}
f\Bigl ( \frac{e}{\sqrt{k}m}\Bigr )\equiv \frac{r_0^2}{r_e^2}=1-\Bigl (\frac{e}{\sqrt{k}m}\Bigr )^{-2}=0,
\label{cond-fin}
\end{eqnarray}
where $r_e$ and $r_0$ are defined by Eqs.~$(\ref{re})$ and $(\ref{r0})$, respectively.
The solution of this equation with the positive value of the mass $m=m_{cl}$,  where
\begin{eqnarray}
m_{cl} =\sqrt{\frac{e^2}{k}},
\label{class-mass}
\end{eqnarray}
gives the mass of the system called  ``the point-charge". 
In classical electrodynamics, the electric charge can be arbitrary, therefore the typical length of the system $r_e$ given by Eq.~$(\ref{re})$ can  be large 
for large $|e|$ 
and the system can be macroscopic.  Having in mind to consider elementary particles, we put $|e|=4.80 \cdot 10^{-10}$ esu for which 
$r_e=1.38 \cdot 10^{-34}$ cm. It is  this system which will be referred to as ``the classical electron".
Its mass is equal to $1.86 \cdot 10^{-6}$ g that is much larger than the experimental value of the real electron mass. 

If  condition $(\ref{cond-fin})$ is 
fulfilled, then according to 
definitions  
$(\ref{rg}-\ref{r0})$  and $(\ref{rho-min})$ 
\begin{eqnarray}
r_0=0,\;\;\;r_e=r_g/2,\;\;\;\rho_{min}=0.
\label{r0-zero}
\end{eqnarray}
Note that owing to the singularity at $\rho=0$ the limit $m \to m_{cl}$ for fixed $e$ is not trivial. Indeed, at  $m = m_{cl}$ we have
$r_0=0$ and the total lagrangian density is zero according to Eq.~$(\ref{ltot-nd})$. Nevertheless the total lagrangian being the integral of the lagrangian 
 density over the three-dimensional space  is nonzero according to Eq.~$(\ref{ltot-final})$: 
$L_{tot}=2m_{cl}c^2$.

Using   Eqs.~$(\ref{r0-zero})$
we get for  $\mathcal{D}(\rho)$,  $\mathcal{N}(\rho)$, $\mathcal{D}'(\rho)$, $\mathcal{N}'(\rho)$,  from 
Eqs.~$(\ref{def-D})$ and $(\ref{def-N})$
\begin{eqnarray}
\mathcal{D}(\rho) =1+\frac{r_g}{2\rho}=1+\frac{r_e}{\rho},
\label{def-D-1}\\
\mathcal{N}(\rho) =1,
\label{def-N-1}\\
\mathcal{D}'(\rho) =-\frac{r_g}{2\rho^2}=-\frac{r_e}{\rho^2},
\label{def-D1-1}\\
\mathcal{N}'(\rho) \equiv 0.
\label{def-N1-1}
\end{eqnarray}
Due to Eq.~$(\ref{def-N1-1})$ formula $(\ref{Wmu-lamu})$ is simplified: 
\begin{eqnarray}
U_{\mu}^{\;\;\lambda \mu}=-U_{\mu}^{\;\;\mu \lambda}\equiv 0,
\label{Umu-lamu}
\end{eqnarray}
while formula $(\ref{W0-0lam})$ becomes
\begin{eqnarray}
U_{0}^{\;\;0 \lambda}=-U_{0}^{\;\; \lambda 0}=\frac{m_{cl}c^2}{4 \pi}
\frac{n_{\lambda}} {\rho^2(1+r_e/\rho)}
\label{U0-0lam} 
\end{eqnarray}
if  Eqs.~ (\ref{def-D-1}--\ref{def-D1-1}),  (\ref{rg}--\ref{re}), and $(\ref{def-kappa})$ are taken into account.

Formula  $(\ref{Pi-Uikl})$ for the total energy of the system of the electromagnetic and gravitational fields reads now
 \begin{eqnarray}
\mathcal{E}\equiv c P_0=\int_{\Omega_{\mathcal{R}}}U_{0}^{\;\;0 \lambda}n_{\lambda}d \sigma
-\int_{\Omega_{\epsilon}}U_{0}^{\;\;0 \lambda}n_{\lambda}d \sigma
\label{E-tot-1}
\end{eqnarray} 
with $\mathcal{R} \to \infty$ and $\epsilon \to 0$. The first integral over the sphere with the radius $\mathcal{R}$   in the right-hand side of  this formula is
 \begin{eqnarray}
\lim_{\mathcal{R} \to \infty} \int \frac{m_{cl}c^2}{4 \pi}
\frac{n_{\lambda}^2} {\mathcal{R}^2(1+r_e/\mathcal{R})} \mathcal{R}^2 d \Omega =m_{cl}c^2
\label{E-tot-first}
\end{eqnarray} 
where $d \sigma=\mathcal{R}^2  d \Omega$ with $ d \Omega=\sin \theta d \theta d \varphi$ being the differential of the solid angle. The second integral in
Eq.~$(\ref{E-tot-1})$ over the sphere with the infinitesimal radius $\epsilon$ is zero. Indeed, we have
 \begin{eqnarray}
\lim_{\epsilon \to 0} \int \frac{m_{cl}c^2}{4 \pi}
\frac{n_{\lambda}^2} {\epsilon^2(1+r_e/\epsilon)} \epsilon^2 d \Omega =0.
\label{E-tot-sec}
\end{eqnarray} 
This means that in spite of the singular behaviour of the electromagnetic and gravitational fields near $\rho=0$, the singularity does not contribute to  the 
surface integral, therefore the energy is determined by the field behaviour at large distances 
($\mathcal{R} \to \infty$).
The net result  of Eqs.~(\ref{E-tot-1}--\ref{E-tot-sec})  is 
 \begin{eqnarray}
\mathcal{E} = m_{cl}c^2=c^2\sqrt{\frac{e^2}{k}}
\label{E-mc2}
\end{eqnarray} 
which shows that the total energy of the system  of the electromagnetic and gravitational fields is equal to $c^2\sqrt{e^2/k}$ or equivalently to $mc^2$ 
with $m=m_{cl}$ according to Eq.~$(\ref{cond-fin})$, where $m$ is the parameter of the RN
 solution. This parameter is now the inertial mass of the system of the electromagnetic 
 and gravitational fields. In principle, the point-like particle with a bare mass $m_b$ may contribute to the total inertial mass of the electron to make it equal to 
the gravitational mass. 
As will be shown in  the next section $m_b=0$.

The total energy-momentum  pseudo-tensor is described with formulas $(\ref{Tik-U})$,   $(\ref{Umu-lamu})$, and  
$(\ref{U0-0lam})$ which give the only nonzero component for the case under consideration
 \begin{eqnarray}
\mathcal{T}_{0}^{0}=\frac{\partial}{\partial \rho^{\lambda}}U_{0}^{\;\;0 \lambda}=
\frac{m_{cl}c^2}{4\pi}\Bigl [\frac{4\pi \rho\delta({\bf R})}{\rho+r_e}
+ \frac{r_e(n_{\lambda})^2}{\rho^4(1+r_e/\rho)^2}
\Bigr ],
\label{tau-0_0}
\end{eqnarray} 
where $\delta({\bf R})$  
denotes the three-dimensional Dirac delta function, and ${\bf R}=(\rho_x, \rho_y,\rho_z)$ is the three-vector. Since 
$\rho\delta({\bf R})=0$ 
the first term in the square brackets in Eq.~$(\ref{tau-0_0})$ does not contribute to $\mathcal{T}_{0}^{0}$, hence the final result is
 \begin{eqnarray}
\mathcal{T}_{0}^{0}= \frac{km_{cl}^2}{4\pi\rho^4(1+r_e/\rho)^2}=\frac{e^2}{4\pi\rho^4(1+r_e/\rho)^2} ,
\label{tau-0_0-fin}
\end{eqnarray} 
where in the right-hand side of Eq.~$(\ref{tau-0_0-fin})$ condition $(\ref{cond-fin})$ is taken into account.
We see that the energy of  the electromagnetic  and gravitational fields for  the classical electron  is localized in the space region of the range of  about 
$r_e=1.38 \cdot 10^{-34}$ cm. Strictly speaking it is true  only for the convenient coordinate system under consideration since the energy distribution cannot be 
uniquely 
defined in field theory \cite{LL,LDF} even if the time variable is fixed.
Nevertheless, $r_e$ can really characterize the order of magnitude of the system length. This is easily seen from the formula 
$(\ref{ltot-final})$ for the total lagrangian which can be rewritten with the help of Eq.~$(\ref{re})$ for the classical electron in the form
 \begin{eqnarray}
L_{tot}= 2c^2\sqrt{\frac{e^2}{k}}=2\frac{e^2}{r_e}.
\label{re-ltot}
\end{eqnarray} 

Let us again calculate  the total energy of the classical electron with the help of Eqs.~$(\ref{Pi-Tik})$ and $(\ref{tau-0_0-fin})$
rather than Eqs.~$(\ref{Pi-Uikl})$ and $(\ref{W0-0lam})$.
 Since now $\rho_{min}=0$ we have
 \begin{eqnarray}
\mathcal{E} =\int_{0}^{\infty}
\mathcal{T}_{0}^{0}(4\pi\rho^2 d \rho)=\int_{0}^{\infty}\frac{e^2 (4\pi\rho^2 d \rho)}{4\pi\rho^4(1+r_e/\rho)^2}.
\label{E-rho-int}
\end{eqnarray}
As seen from this formula the integrand increases with decreasing of $\rho$  as $\rho^{-2}$ at  $\rho \gg r_e$ but
at $\rho \leq r_e$ the integrand  goes to a  finite constant. As a result, the integral is convergent and is equal to
$m_{cl}c^2=c^2\sqrt{e^2/k}$. Indeed, using Eq.~$(\ref{re})$ we have from  Eq.~$(\ref{E-rho-int})$
 \begin{eqnarray}
\mathcal{E} =\int_{0}^{\infty}\frac{e^2}{(\rho+r_e)^2}  d \rho=\frac{e^2}{r_e}=c^2\sqrt{\frac{e^2}{k}}=m_{cl}c^2.
\label{E-rho-int-1}
\end{eqnarray}
In the absence of gravitation, when $k=0$, the value of $r_e$ is zero according to Eq.~$(\ref{re})$ and the integral in Eq.~$(\ref{E-rho-int})$ becomes divergent as in
classical electrodynamics. Therefore it is the gravitational interaction which makes the classical electron mass finite. Roughly speaking the gravitation
contribution reduces the electron mass from infinity (the pure electromagnetic mass) to a finite mass
$\sqrt{e^2/k}$.

It is interesting to compare formula $(\ref{tau-0_0-fin})$ with the energy-momentum tensor density  of the electromagnetic field.
Since the tensor component $T_0^0$ is the same for coordinate systems $x^0$, $r$, $\theta$, $\phi$  and $\rho_0$, $\rho_x$, 
$\rho_y$, $\rho_z$, the expression for the tensor density component $\tilde{T}_0^0$ of the electromagnetic fields 
follows from  Eqs.~$(\ref{T-00})$,  $(\ref{r-rho})$, $(\ref{det-h-nd})$, $(\ref{def-D-1})$, and $(\ref{def-N-1})$
 \begin{eqnarray}
\tilde{T}_{0}^{0} \equiv |h| T_0^0=\frac{e^2 N(\rho)[D(\rho)]^2 }{8\pi[\rho D(\rho)]^4}
=\frac{e^2}{8\pi\rho^4(1+\frac{r_e}{\rho})^2} .
\label{T-em-0_0-den}
\end{eqnarray} 
A comparison of Eqs.~$(\ref{tau-0_0-fin})$ and  $(\ref{T-em-0_0-den})$ shows that the total energy-momentum pseudo-tensor
 density is by a factor of two larger than that of the electromagnetic field. This is the trivial consequence of the Tolman formula \cite{Tolm}
  \begin{eqnarray}
\mathcal{T}_{0}^{0}=  
\tilde{T}_{0}^{0}-  \tilde{T}_{1}^{1}-\tilde{T}_{2}^{2}-\tilde{T}_{3}^{3}.
\label{Tolman}
\end{eqnarray}  
Indeed, the trace of the tensor  density $\tilde{T}_{i}^{i}$ is zero for the electromagnetic field \cite{JDJ,LL,SW}, hence
$-\tilde{T}_{1}^{1}-\tilde{T}_{2}^{2}-\tilde{T}_{3}^{3}=\tilde{T}_{0}^{0}$, therefore $\mathcal{T}_{0}^{0}=2\tilde{T}_{0}^{0}$ which proves the statement. Since 
the 
density of the energy-momentum tensor
of the electromagnetic field  $\tilde{T}_{0}^{0}$ is one-half of the total pseudo-tensor density $\mathcal{T}_{0}^{0}$, the electromagnetic mass of the classical
electron is also  one-half of the total mass $m_{cl}$ according to Eq.~$(\ref{Pi-Tik})$. Note that ``the electromagnetic mass" has conditional meaning since its 
formula contains the gravitational constant $k$ according to Eqs. $m_{em}=m_{cl}/2$ and $(\ref{class-mass})$.

The space part of the energy-momentum pseudo-tensor density  $\mathcal{T}_{\mu}^{\nu}$ is zero
 \begin{eqnarray}
 \mathcal{T}_{\mu}^{\nu} \equiv 0,
\label{tau-mu-nu-zero}
\end{eqnarray} 
which follows from Eqs.~$(\ref{Tik-U})$ and $(\ref{Umu-lamu})$. This assumes the absence of any pressure of any part of the system
under discussion  (classical electron) on others. Therefore there is no need in the additional surface-tension of the charged liquid (existing,
 for instance, in the Lorentz model of the electron \cite{LOR}) which
prevents disintegration of the classical electron.  There is an equilibrium between the electrostatic repulsion and gravitational attraction.

Since the components $U_{\mu}^{\;\;0 \lambda}$ of the superpotential are zero the three-momentum of the classical electron is zero 
in the rest system of frame. If the four-velocity of the singular point (the point with $\rho=0$ in the rest system) is $u_i$ than 
the four-vector of the electron in any Lorentz system is
 \begin{eqnarray}
P_i=m_{cl}cu_i,
\label{P-i=mc*u-i}
\end{eqnarray}
while the components of the energy-momentum pseudo-tensor are
 \begin{eqnarray}
\mathcal{T}_{i}^{k}=\frac{e^2}{4\pi\rho^4(1+r_e/\rho)^2} u_i u^k.
\label{Tik-ui-*u-k}
\end{eqnarray}
Since the function of $\rho$ in the right-hand side of Eq.~$(\ref{Tik-ui-*u-k})$ has a sharp peak  near $\rho=0$ and the integral in 
 Eq.~$(\ref{E-rho-int})$ is equal to $m_{cl}c^2$ according to formulas $(\ref{E-rho-int})$ and  $(\ref{E-rho-int-1})$ it is possible to approximate 
$\mathcal{T}_{i}^{k}$
 with the relation
 \begin{eqnarray}
\mathcal{T}_{i}^{k}=m_{cl}c^2 \delta({\bf R}) u_i u^k
\label{Tik-point}
\end{eqnarray}
using the Dirac three-dimensional $\delta$-function. This formula is widely used to describe  point-like particles. 

We would like to stress that 
Eq.~$(\ref{Tik-point})$ takes into account the electromagnetic field of the classical electron, hence it is not correct to add to the tensor density given by 
Eq.~$(\ref{Tik-point})$ the energy-momentum tensor density of either the external or the total electromagnetic field as it is often done in classical electrodynamics 
(see 
for instance \cite{LL,SW}).  Let us explain this statement in more details. If the total    electromagnetic  field tensor $F_{(tot)}^{lm}$ is equal to the sum of the external field tensor  $F_{(ext)}^{lm}$ and that of
 the classical electron field $F_{(cl)}^{lm}$ , then the total energy-momentum tensor density of the electromagnetic field is
 \begin{eqnarray}
\tilde{T}_{{(tot)}\;i}^{k}=|h|\{T_{{(cl)}\;i}^{k}+T_{{(ext)}\;i}^{k}+T_{{(int)}\;i}^{k}\}.
\label{Tik-tot}
\end{eqnarray}
 Here,    $T_{{(cl)}\;i}^{k}$ and    $T_{{(ext)}\;i}^{k}$ are respectively the  energy-momentum tensors of the classical electron and the external field given by    
Eq.~$(\ref{Tik})$  for $F^{lm}=F_{(cl)}^{lm}$  and   $F^{lm}=F_{(ext)}^{lm}$, respectively. Here $F_{(cl)}^{kl}$ and $F_{(ext)}^{kl}$
are the tensors of the electromagnetic fields of the electron and the external field.  
The interference term   $T_{{(int)}\;i}^{k}$   is
\begin{eqnarray}
T_{(int)i}^k=\frac{1}{4 \pi}\Bigl \{ \frac{1}{2}\delta^{k}_{i} F_{(cl)lm}F_{(ext)}^{lm}
 -F_{(cl)}^{kl}F_{(ext)il}- F_{(ext)}^{kl}F_{(cl)il} 
\Bigr \}.
\label{Tik-int} 
\end{eqnarray}
In the zero approximation applied here, we ignore the alteration of the metrics due to the influence of the external electromagnetic field.
Since the contribution of 
$T_{{(cl)}\;i}^{k}$ is taken into account in the term given by Eq.~$(\ref{Tik-ui-*u-k})$ or $(\ref{Tik-point})$ the additional terms describing the contribution of the 
external electromagnetic field is given by the formula
 \begin{eqnarray}
\Delta\tilde{T}_{{(tot)}\;i}^{k}=|h|\{T_{{(ext)}\;i}^{k}+T_{{(int)}\;i}^{k}\}
\label{del-Tik-tot}
\end{eqnarray}
rather than $|h| T_{{(ext)}\;i}^{k}$ or $|h| T_{{(tot)}\;i}^{k}$. The expression for $\Delta\tilde{T}_{{(tot)}\;i}^{k}$ in Eq.~$(\ref{del-Tik-tot})$ is integrable 
function while  $|h| T_{{(tot)}\;i}^{k}$ is not.
\section{Equivalence principle}
\label{sect-8}
In order to  calculate the total gravitational mass
$m_{gr}$, of any insular system we should consider the $g_{00}$ component of the metric tensor which asymptotic behaviour at $\rho \to \infty$ is
\cite{LL,SW}
\begin{eqnarray}
g_{00} \approx 1+2\frac{\phi(\rho)}{c^2}=1-\frac{2k m_{gr}}{c^2 \rho},
\label{g00-asy}
\end{eqnarray} 
where $\phi(\rho)$ denotes the Newtonian gravitational potential. Using formulas 
$(\ref{def-D})$   and     $(\ref{def-N})$   
respectively for $\mathcal{D}$ and  $\mathcal{N}$ 
we get from Eq.~$(\ref{ds2-xyz})$ that for any $e$ and $m$ at $\rho \to \infty$
\begin{eqnarray}
g_{00} =\frac{\mathcal{N}^2}{\mathcal{D}^2} \approx 1-\frac{r_g}{\rho}=1-\frac{2k m}{c^2 \rho}.
\label{g00-asy-nd}
\end{eqnarray} 
A comparison of Eq.~$(\ref{g00-asy})$ with $(\ref{g00-asy-nd})$ gives $m_{gr}=m$ which means that the parameter $m$ in the  RN
 solution is always the total gravitational mass of the system. 
For  $m=m_{cl}=\sqrt{e^2/k}$, the parameter $m$ becomes  the inertial mass of the system of the electromagnetic and gravitational fields in accordance with 
Eqs.~$(\ref{E-tot-1})$ and $(\ref{E-mc2})$.  Therefore for this case,  the total gravitational mass is equal to the inertial mass of the system of the electromagnetic 
and gravitational fields.   According to the equivalence principle, this means that the bare mass $m_{b}$ of the point-like particle is zero. 

Therefore the classical electron is a system of the electromagnetic and gravitational fields localized in the space region with the typical length of about 
$10^{-34}$ cm. There is no need in the existence of any charged point-like particle which is usually  named ``electron" in classical electrodynamics.
\\$\;\;\;\;$\\

\section{Electrical charge distribution}
\label{sect-9}
In order to study the electrical charge distribution, we should consider the non-trivial properties of the three-dimensional space in the vicinity of the 
point with $\rho=0$. Let $\rho$ be equal to infinitesimal $ \epsilon>0$.
 In order to calculate the length of circumference of a circle with the maximal length on the sphere with  $\sqrt{\rho_x^2+\rho_y^2+\rho_z^2}=\epsilon$ we should take 
into account 
Eq.~$(\ref{def-ds2})$ and put $\theta$ equal to $\pi /2$. We have the formula for the length
 \begin{eqnarray}
l_{\epsilon} =\int_{0}^{2\pi}D(\epsilon) \epsilon d \phi=2\pi(\epsilon+r_e).
\label{length}
\end{eqnarray} 
In the transformation of the right-hand side of this equation, expression $(\ref{def-D-1})$ for $D(\rho)$ is used. Formula $(\ref{length})$ shows that for $\epsilon=0$, one gets a nonzero length $l_0=2\pi r_e$. 

In order to understand this paradox, we find the relation between $r$ and $\rho$, which follows from  Eqs.~$(\ref{r-rho})$ and $(\ref{def-D-1})$
\begin{eqnarray}
r=\rho+r_e.
\label{r-rho1}
\end{eqnarray}
As seen from  Eqs.~$(\ref{r-rho1})$  the point $\rho=0$ in the space $(\rho_x,\rho_y,\rho_z)$ corresponds to the sphere
with the radius $r_e$ in the three-dimensional space $(x,y,z)$, where
\begin{eqnarray}
x=r\sin \theta \cos \phi,\;\;\; y=r\sin \theta \sin \phi,\;\;\; z=r\cos \theta. 
\label{xyz-1}
\end{eqnarray} 
The electric charge density on this sphere is
\begin{eqnarray}
\rho_{ch}=\frac{e}{4\pi r_e^2}=\frac{c^4}{4\pi e k}
\label{ro_ch}
\end{eqnarray}
since the metric on this sphere in accordance with  Eqs.~$(\ref{g22})$ and  $(\ref{g33})$  is the same as on  the sphere in the Euclidean space.

The radial distance between points with $\rho=\rho_1>0$ and $\rho=\rho_2>0$, the other coordinates being equal, is
\begin{eqnarray}
\nonumber
l_{12}=\int_{\rho_1}^{\rho_2}D(\rho)d \rho=\int_{\rho_1}^{\rho_2}\Bigl (1+\frac{r_e}{\rho}\Bigr )d \rho = \\
=\rho_2-\rho_1+r_e\ln\Bigl (\frac{\rho_2}{\rho_1} \Bigr )
\label{l12}
\end{eqnarray}
 in accordance with the metrics given by Eq.~$(\ref{def-ds2})$ and expression $(\ref{def-D-1})$ for $D(\rho)$. As is seen from the above formula 
$l_{12} \to \infty$ if  $\rho_1 \to 0$. Therefore the distance between the point with $\rho_x=\rho_y=\rho_z=0$ and any other point is infinite. 
Since for $m=m_{cl}$ we have $r_g=2r_e$ and the tensor component  $-g_{11}$ for the coordinates $r$, $\theta$, and $\varphi$
is equal to $1/(1-r_e/r)^2$ according to Eqs.~$(\ref{g00-fin})$ and $(\ref{lamb-1})$. Therefore $-g_{11}$ goes to infinity when $r \to r_e$.
It is obvious from this behavior of $-g_{11}$ 
that the distance between any point in the $(x,y,z)$ space with a finite $r>r_e$ and the sphere with the radius $r_e$,
which center has coordinates $x=y=z=0$,  is also infinite. For $m=m_{cl}$  $g_{00}=-1/g_{11}=(1-r_e/r)^2$ according to Eqs.~$(\ref{g00-fin})$ and 
$(\ref{lamb-1})$. The sphere with $r=r_e$ represents the surface of the event horizon ($g_{00}(r_e)=0$), therefore no information can be obtained from the internal 
($r<r_e$) region  
\cite{LL,SW}.
This means that  the coordinates $x$, $y$, $z$ do not correspond to any physical objects which can be observed by any external observer
 if $\sqrt{x^2+y^2+z^2}<r_e.$ This means also that any observables (energy, electric and magnetic field strengths etc.) can be really measured
only in the three-dimensional space with $r>r_e$. This explains why the integral in Eq.~(\ref{E-rho-int-1}) runs only for $r\geq r_e$ which corresponds to $\rho \geq 
0$.

Also, the electric charge of the classical electron is uniformly distributed on the sphere with $r=r_e$.  The space-like components of 
the energy-momentum pseudo-tensor density $\mathcal{T}_{\mu}^{\nu} $
 responsible for the forces between parts of the electron are identically zero. Therefore the problem of extra nonelectromagnetic forces preventing the electron 
 disintegration  as in the Lorentz model \cite{LOR}, presenting it as a charged liquid drop, is absent.    The distance from this charged sphere and any point in the 
three-dimensional space $(x,y,z)$ is infinite.
\section{Discussion of results and quantum effects}
\label{sect-10}
As is shown above the system of gravitational and electromagnetic fields described by Eqs.~(\ref{g00-fin}--\ref{g33}) and 
$(\ref{sol-F01})$   has a finite total inertial mass if the parameter $m$ of 
the RN solution is equal to  $\sqrt{e^2/k}$. For this case $m$ is both the total inertial and gravitational mass,
that is in agreement with the equivalence principle. Such a system is called  ``the classical electron" when $e$ is the experimental electron charge.
There is no need in any additional charged point-like particle which is usually called the electron. 
This assumes the absence of the term in the action describing the interaction
of the point-charge moving with a three-velocity ${\bf v}$ with the electromagnetic field usually considered in classical electrodynamics \cite{JDJ,LL,RPF}
\begin{eqnarray}
S_{pf}=e\int \Bigl [\frac{{\bf v}(t)}{c}{\bf A}(x,y,z,t)-A_0(x,y,z,t)\Bigr ]dt.
\label{lagr-point}
\end{eqnarray}  
Note that the term $S_{pf}$ has no physical meaning since it is infinite. Indeed, we should consider the scalar $A_0$ and vector ${\bf A}$ 
potentials of the total electromagnetic field.  But  $A_0(x,y,z,t)$ and ${\bf A}(x,y,z,t)$ are infinite at any point of a trajectory of the point-charge. 
Consideration of the external  field,  usually done in textbooks for the integral in Eq.~$(\ref{lagr-point})$ to  make  it convergent, is in 
 contradiction with the fundamental concepts of the field theory.  Indeed, all the basic quantities should be expressed in terms of variables of the total
electromagnetic field as it is the only fundamental notion. The term ``external field" for all point-charges assumes the consideration of individual fields of all 
point-charges but these fields are not fundamental notions.

In the  approach under consideration in the present paper, the only existing entities are the electromagnetic and gravitational fields, while 
the  solutions, localized in the space regions of the range of about   $10^{-34}$ cm,
represent  the classical electrons. The important thing is the tetrad representation, which makes the action finite. 
Considering the metric tensor components as the true variables of the gravitational field we conclude that the action for the classical electron is infinite.

It is well known that the quantum effects become important  at  distances of about the Compton wavelength of the real electron 
$\lambda _C=\hbar/(m_e c)$  where $\hbar$ denotes the Plank constant and $m_e$ denotes the experimental electron mass. Since $\lambda _C=3.86\cdot 10^{-11}$ cm,   it 
is much greater than the typical length $r_e =1.38 \cdot 10^{-34}$ cm  for the classical electron.  For the first sight, this means that the above consideration 
has no physical meaning. Nevertheless, highly  likely that this argument is not true. Indeed, the divergence of the integral $\int T^0_0d^3r$ in classical 
electrodynamics is very hard, namely the integrand behaves as $1/r^2$ at $r \to 0$. 
Nevertheless the contribution of gravitation makes the energy finite for the parameter  $m$ of the Reissner--Nordstr\"om solution 
equal to $\sqrt{e^2/k}$.
Since the divergence of the Feynman graphs describing the electromagnetic field contribution to the electron mass is only logarithmic in QED (the integrand behaves as 
$1/r$
at small $r$), the role of gravitation can be much less important. If it changes the behaviour of the integrand, say, to $1/r^{1-\delta}$ with a very  small
positive $\delta$, the integral becomes convergent. The integral for the self-energy in QED is proportional to
$\alpha_{em}\ln[\Lambda_c/(mc)]$ where $\Lambda_c$ is an ultraviolet cutoff parameter which is put equal to infinity and 
$\alpha_{em}=e^2/(\hbar c)$ is the fine-structure constant.  As is seen from Eq.~$(\ref{E-rho-int})$ the space cutoff parameter arising due to the contribution of 
gravitation is $r_e$, hence the natural cutoff parameter in the momentum space is 
\begin{eqnarray}
\Lambda_{c}=\frac{\hbar}{r_e}=\frac{\hbar c^2}{\sqrt{e^2 k}}
=c\sqrt\frac{\hbar c}{e^2} \sqrt{\frac{\hbar c}{k}}
=\frac{cM_{Pl}}{\sqrt{\alpha_{em}}},
\label{cut-gr}
\end{eqnarray}  
where $M_{Pl}= \sqrt{\hbar c/k}$ is the Plank mass. Therefore $\alpha_{em}\ln(\Lambda_c/(mc))$ could be replaced by $\alpha_{em}\ln(M_{Pl}/m)$ with the 
same accuracy. In the classical physics, the finite energy of the electron exists for $m=\sqrt{e^2/k}$ only or, in  other words, when the function 
$f(e/m\sqrt{k})$ defined by Eq.~$(\ref{cond-fin})$ is zero. It is not excluded, that  for the quantum case the finite energy of the field configuration exists 
not for all mass parameters $m$ but for some fixed one for which the quantum analog of the function $f(x)$ is zero with  $x=\alpha_{em}\ln(M_{Pl}/m)$.  For the 
experimental mass of the electron $M_{Pl}/m_e \approx 2.4\cdot 10^{22}$, while  $\alpha_{em}=1/137.04$, therefore $x \approx 0.38$. This means that $x$ is of the 
order of unity, hence the argument  $x=\alpha_{em}\ln(M_{Pl}/m)$ looks rather natural and equation $f(x)=0$ could give the electron mass close to its experimental 
value, while, for instance, $x=\hbar/(r_em_ec)$ is unnatural (huge). 
    
The most natural question arises: "How to find the quantum analog of the function $f(x)$ defining the electron mass?" We saw that the total stress tensor 
$\mathcal{T}^{\nu}_{\mu}$ is  identically zero in Eq.~$(\ref{tau-mu-nu-zero})$ if $m=\sqrt{e^2/k}$ and vice versa. It is likely that in the quantum case this 
condition can be replaced with the demand that some matrix  elements of the total stress tensor over the wavefunctions of the electron must be zero to prevent 
desintegration of the electron by the electromagnetic forces. And this demand defines the quantum analog of the function $f(x)$. 

By changing the sign of the charge in Eq.~$(\ref{sol-F01})$ we get a set of the electromagnetic and 
gravitational fields corresponding to  ``classical positron". 
 It has the same  mass as the electron in the approach under discussion, hence  the positron is not an electron with the negative energy as in the Dirac 
equation \cite{RPF}. It is assumed that in the future version of the quantum field theory, there will be no need in the local 
electron-positron field, that has the negative vacuum energy. 

In the approach based on the supersymmetry, the negative vacuum energy of the 
electron-positron field compensates the positive vacuum energy of the photon field. The total vacuum energy of all
 fundamental  fields can be equal to zero if the supersymmetry is not broken. 
Supersymmetric partners of the existing particles with masses less than about 1 
TeV/c$^2$ are not found up to now (see for instance \cite{no-susy-t,no-susy-a,no-susy-c}). This can mean that the supersymmetry is not a fundamental symmetry of  elementary particles. In the absence  of the electron-positron field, the vacuum energy 
can probably be made finite due to the negative contribution  of the  gravitational field. Indeed, let us consider only the modes  of the quantum oscillations
 of the electromagnetic field with frequencies $\omega < \omega_c$. Let us imagine that the vacuum energy density for these modes 
with the mode energies $\epsilon_{vac}=\hbar \omega/2$ is huge but finite in the absence of the gravitational interaction. If 
the gravitational interaction is switched on its 
negative contribution decreases the vacuum energy. If the electromagnetic energy density goes to infinity (when $\omega_c \to \infty$) as in standard QED the modulus 
of the  gravitation  interaction  contribution  increases also to infinity. It is not excluded that the total vacuum energy density is finite. This picture is in an 
analogy with the classical 
 electron case for which the infinite  growth of the electromagnetic energy density at $\rho \to 0$ is restricted with the gravitational attraction of small space 
regions, filled with electromagnetic field, with each other. The cancellation of the contributions to the total mass of the electromagnetic and the gravitational 
fields leads to the finite value of the total mass.
  
Electrons take part in weak interaction and this electron property should be taken into account. Therefore we should try to find solutions of equations for the system 
of the electromagnetic, gravitational,  and  weak-boson fields.  Another property
of the electron, that should be taken into account in its realistic description, is the electron spin $s$ equal to $\hbar/2$.
 But   solutions with $s=\hbar/2$  are not excluded for the nonlinear boson fields. A well known example provides the Skyrme model \cite{Sk,Sky} in which  the 
solutions with the spin $\hbar/2$ (baryons) are constructed though the fundamental field is the nonlinear field of the 
pseudoscalar pions. Another way to make the spin is to consider a solution (if it exists) which corresponds to an electric charge and a magnetic dipole moment since 
the electromagnetic 
field in this case has an angular momentum \cite{PAMD}. The localized states of the quantized electromagnetic, gravitational,  and  weak-boson fields with the 
$\hbar/2$ spin, the observed values of the electric 
charge,  the weak  charge, and the finite masses equal to those of the electron, muon, and $\tau$-lepton could exist. In the same way, the localized 
solutions of the quantized gluon, electromagnetic, weak boson, and gravitational field equations would be quarks and there would not be a 
need in local bispinor fields corresponding to the point-like massive quarks. This problem cannot probably be solved soon since there is  no renormalizable  quantum 
field theory of 
gravitation though any estimates of the electron mass could be performed in lattice calculations using the continual integrals for electromagnetic and gravitational 
fields (see review \cite{FP}). Nevertheless, we assume that the idea to construct all observed particles  as  singular or localized (in a very small 
three-dimensional space region) solutions of fundamental field equations is constructive.
\section{Conclusions}
\label{sect-11}
It is shown, that for the Reissner--Nordstr\"om solution,
the contribution of gravitation  makes the total energy-momentum preudo-tensor density integrable function if the parameter of the solution $m=\sqrt{e^2/k}$. 
Nevertheless the 
singular point exists for this case also. The total inertial mass of the system of the electromagnetic and gravitational fields is finite ($m_{in}=\sqrt{e^2/k}$) 
and equal to its total gravitational mass $m_{gr}$. 
According to
the equivalence principle ($m_{in}=m_{gr}$),
 this means the absence of an additional contribution to the total mass of any charged point-like particle with a nonzero bare mass. 
In the approach of the present paper, the classical electron is the system of the electromagnetic and 
gravitational fields localized in the space region with the typical length of about $10^{-34}$ cm,  $e=-4.80 \cdot 10^{-10}$ esu,
and $m=1.86 \cdot 10^{-6}$~g.

Since the total 
stress tensor density $\mathcal{T}_{\mu}^{\nu}$  is identically zero there is no need in additional  nonelectrostatic forces  preventing 
the disintegration of the classical electron which were introduced in the Lorentz model of the electron (surface tension forces for the charged liquid drop). 

As is known the total lagrangian for the Reissner--Nordstr\"om solution has nonintegrable singularity if the metric tensor components are considered as fundamental
variables of the gravitational field. The total lagrangian for the Reissner--Nordstr\"om
solution is shown to be finite in the tetrad representation for any values of the parameters $e$ and $m$.

We assume that it is not excluded that in quantum field theory, the only existing fundamental entities are gravitational, electromagnetic, weak-boson, and gluon 
fields. Leptons and quarks are states of these fields having singular points which positions are usually considered as the positions of corresponding  particles.
\section*{Acknowledgements}
I would like to thank Prof. V.R. Shaginyan for many friendly advices and all participants of Theoretical Physics Division 
and High Energy Physics Division seminars of the Petersburg Nuclear Physics Institute for useful discussions.\\
\section{Appendix}

\subsection{Christoffel symbols in uniform coordinates\\}
\label{appendix-1}

Using the metric tensor defined by  Eq.~$(\ref{ds2-xyz})$ the Christoffel symbols can be obtained with the help of 
Eqs.~$(\ref{Crist-up})$ and $(\ref{Crist-down})$. The nonzero components of $\Gamma_{i,jk}$ in the uniform coordinates are
\begin{eqnarray}
\Gamma_{0,0\mu}=\Gamma_{0,\mu 0}=-\Gamma_{\mu,00}
=\frac{\mathcal{N}n_{\mu}}{\mathcal{D}^3} \{\mathcal{D}\mathcal{N}'-\mathcal{N}\mathcal{D}'\}, 
\label{chri-down-00m}\\
\Gamma_{\mu,\mu\mu}=-n_{\mu}\mathcal{D}\mathcal{D}',
\label{chri-down-mmm}
\end{eqnarray}   
where the three-vector $n_{\mu}$ is defined in Eq.~$(\ref{n_lam})$ and
\begin{eqnarray}
\mathcal{N}'\equiv \frac{\partial \mathcal{N}}{\partial \rho}=-\frac{r_0^2}{2\rho^3},
\label{n-prime}\\
\mathcal{D}'\equiv \frac{\partial \mathcal{D}}{\partial \rho}  =-\frac{r_g}{2 \rho^2}+\frac{r^2_0}{2 \rho^3},
\label{d-prime}\\
\mathcal{D}\mathcal{N}'-\mathcal{N}\mathcal{D}'
=\frac{r_g}{2 \rho^2}\Bigl (1-\frac{r^2_0}{4 \rho^2} \Bigr )-\frac{r^2_0}{\rho^3}.
\label{wronsk-nd}
\end{eqnarray}
Note that in Appendix, there is no summation over two or three identical indexes, all sums contain the symbol $\sum$.
The functions $\mathcal{D}$ and $\mathcal{N}$ are defined in Eqs.~$(\ref{def-D})$ and $(\ref{def-N})$, while 
$r_g$ and  $r_0$ 
in Eqs.~$(\ref{rg})$ and $(\ref{r0})$, respectively. 
Other nonzero Christoffel symbols for $\mu \neq \nu$ are
\begin{eqnarray}
\Gamma_{\mu,\nu\nu}=-\Gamma_{\nu,\mu\nu}=-\Gamma_{\nu,\nu\mu}=n_{\mu}\mathcal{D}\mathcal{D}'. 
\label{chri-down-mnn}
\end{eqnarray}

The nonzero Christoffel symbols $\Gamma^i_{jk}$ are
\begin{eqnarray}
\Gamma^{0}_{0\mu}=\Gamma^{0}_{\mu 0}=n_{\mu}\frac{\{\mathcal{D}\mathcal{N}'-\mathcal{N}\mathcal{D}'\}}{\mathcal{N}\mathcal{D}},
\label{chri-up-00m}\\
\Gamma^{\mu}_{00}=n_{\mu}\mathcal{N}\frac{\{\mathcal{D}\mathcal{N}'-\mathcal{N}\mathcal{D}'\}}{\mathcal{D}^5},
\label{chri-up-m00}\\
\Gamma^{\mu}_{\mu\mu}=n_{\mu}\frac{\mathcal{D}'}{\mathcal{D}},
\label{chri-up-mmm}
\end{eqnarray}
while for $\mu \neq \nu$
\begin{eqnarray}
\Gamma^{\mu}_{\nu\nu}=-\Gamma^{\nu}_{\mu\nu}=-\Gamma^{\nu}_{\nu\mu}=-n_{\mu}\frac{\mathcal{D}'}{\mathcal{D}}.
\label{chri-up-mnn}
\end{eqnarray}
\subsection{Covariant derivatives of tetrads\\} 
\label{appendix-2}

For the tetrads defined by Eqs.~$(\ref{rn-tetr})$ and the Christoffel symbols obtained in 
Subsec. \ref{appendix-1}
the calculation of the covariant derivatives of $h^{(a)}_{n}$
with using Eq.~$(\ref{cov-der})$ gives the following nonzero components:
\begin{eqnarray}
h^{(0)}_{\mu; 0}=-n_{\mu}\frac{\{\mathcal{D}\mathcal{N}'-\mathcal{N}\mathcal{D}'\}}{\mathcal{D}^2},
\label{h0-0m}\\
h^{(\mu)}_{0; 0}=-n_{\mu}\frac{\mathcal{N}}{\mathcal{D}^4}\{\mathcal{D}\mathcal{N}'-\mathcal{N}\mathcal{D}'\},
\label{hm-00}   
\end{eqnarray}
and for $\mu \neq \nu$ 
\begin{eqnarray}
h^{(\mu)}_{\nu; \nu}=n_{\mu}\mathcal{D}',
\label{hm-nn} \\
h^{(\nu)}_{\mu; \nu}=-n_{\mu}\mathcal{D}'.
\label{hn-mn} 
\end{eqnarray}

As a consequence of these formulas, we have
\begin{eqnarray}
\sum_{s=0}^3h^{(0)s}_{\;\;\;\;\;;s}=0,
\label{h0-summ}\\
\sum_{s=0}^3h^{(\mu)s}_{\;\;\;\;\;;s}=-\frac{n_{\mu}}{\mathcal{D}}\Bigl \{ 
\frac{\mathcal{N}'}{\mathcal{N}}+\frac{\mathcal{D}'}{\mathcal{D}}\Bigr \},
\label{hmu-summ}   
\end{eqnarray}
therefore 
\begin{eqnarray}
\sum_{a,b=0}^3\eta_{ab}\sum_{l=0}^3h^{(a)l}_{\;\;\;\;\;;l}
\sum_{s=0}^3h^{(b)s}_{\;\;\;\;\;;s}=-\frac{1}{\mathcal{D}^2}\Bigl \{
\frac{\mathcal{N}'}{\mathcal{N}}+\frac{\mathcal{D}'}{\mathcal{D}}\Bigr \}^2.
\label{ss-2}
\end{eqnarray}
This sum is nothing else than the second sum in the brackets in Eq.~$(\ref{gra-lagr})$.

The first sum in the brackets in Eq.~$(\ref{gra-lagr})$ is
\begin{eqnarray}
\nonumber
\sum_{a,b=0}^3\eta_{ab}\sum_{l,k=0}^3\Bigl \{  h^{(a)l}_{\;\;\;\;\;;k}h^{(b)k}_{\;\;\;\;\;;l}\Bigr \} = 
\sum_{a,b=0}^3\eta_{ab}\sum_{l,k=0}^3\Bigl \{  g^{kk}g^{ll}h^{(a)}_{\;\;\;\;\;l;k}h^{(b)}_{\;\;\;\;\;k;l}\Bigr \}
\\
=-\frac{1}{\mathcal{D}^2} \Bigl \{ \Bigl( \frac{\mathcal{N}'}{\mathcal{N}}\Bigr )^2
-2\frac{\mathcal{N}'}{\mathcal{N}}\frac{\mathcal{D}'}{\mathcal{D}}
+3\Bigl( \frac{\mathcal{D}'}{\mathcal{D}}\Bigr )^2 \Bigr \},
\label{sl-2}
\end{eqnarray}
which follows from Eq.~$(\ref{h0-0m}-\ref{hn-mn})$ for $h^{(a)}_{\;\;\;\;\;l;k}$ and $(\ref{ds2-xyz})$ for $g^{jj}=1/g_{jj}$.
The difference of these two sums multiplied by $|h|/(2\kappa)$ provides  the formula for the
lagrangian density of the gravitational field given by Eq.~$(\ref{lg-nd})$ if formula $(\ref{det-h-nd})$ for the determinant $|h|$ is 
taken into account.
\subsection{Tensor $\gamma_{iml}$ and four-vector $\Phi_m$\\}
\label{appendix-3}

Using definition of $\gamma_{iml}$ given by Eq.~$(\ref{g_ikl})$, 
Eqs.~(\ref{h0-0m}--\ref{hn-mn}) for $ h^{(a)}_{l;k}$ and  
Eqs.~$(\ref{rn-tetr})$ for $h^{(a)}_{m}$ one gets formulas for the nonzero covariant components of  the tensor  $\gamma_{iml}$
\begin{eqnarray}
\gamma_{\mu00}=-\gamma_{0\mu0}=n_{\mu}\frac{\mathcal{N}}{\mathcal{D}^3}
\{\mathcal{D}\mathcal{N}'-\mathcal{N}\mathcal{D}'\},
\label{gam-m00}\\
\gamma_{\nu\mu\nu}=-\gamma_{\mu\nu\nu}=n_{\mu}\mathcal{D}\mathcal{D}',
\label{gam-nmn}
\end{eqnarray}
where $\mu \neq \nu$ in Eq.~$(\ref{gam-nmn})$. 

     Applying the metric tensor from Eq.~$(\ref{ds2-xyz})$ corresponding to the RN
 solution for the 
uniform coordinates the nonzero components of $\gamma^i_{\;\;ml}$ can be obtained
\begin{eqnarray}
\gamma^{\mu}_{\;\;00}=-\gamma_{0\;\;\;0}^{\;\;\mu}=-n_{\mu}\frac{\mathcal{N}}{\mathcal{D}^5}
\{\mathcal{D}\mathcal{N}'-\mathcal{N}\mathcal{D}'\},
\label{gam-m_00}\\
\gamma_{0\mu}^{\;\;\;0}=-\gamma_{\mu 0}^{\;\;\;0}
=n_{\mu}\Bigl [
\frac{\mathcal{D}'}{\mathcal{D}}-\frac{\mathcal{N}'}{\mathcal{N}}\Bigr ],
\label{gam-0m-up-0}\\
\gamma^{0}  _{\;\;\mu0}=-\gamma_{\mu\;\;0} ^{\;\;0} 
=-n_{\mu}\frac{\{\mathcal{D}\mathcal{N}'-\mathcal{N}\mathcal{D}'\}}{\mathcal{N}\mathcal{D}},
\label{gam-0_m0}
\end{eqnarray}
while for $\mu \neq \nu$ we have
\begin{eqnarray}
\gamma^{\nu} _{\;\mu\nu}=-\gamma_{\mu\;\;\nu} ^{\;\;\nu}
=\gamma_{\nu\mu}^{\;\;\;\nu}=-\gamma_{\mu\nu}^{\;\;\;\nu}
=-n_{\mu}\frac{\mathcal{D}'}{\mathcal{D}},
\label{gam-n_mn}\\
\gamma^{\mu}_{\;\;\nu\nu}=-\gamma_{\nu\;\;\;\nu} ^{\;\;\mu} =n_{\mu}\frac{\mathcal{D}'}{\mathcal{D}}.
\label{gam-m_nn}
\end{eqnarray}

In an analogous way, the  formulas
\begin{eqnarray}
\gamma^{\mu 0}_{\;\;\;\;0}=-\gamma^{0\mu} _{\;\;\;\;0}=-\frac{n_{\mu}}{\mathcal{N}\mathcal{D}^3}
\{\mathcal{D}\mathcal{N}'-\mathcal{N}\mathcal{D}'\},
\label{gam-m0-0}
\end{eqnarray}  
and the relations for $\mu \neq \nu$ 
\begin{eqnarray}
\gamma^{\nu \mu} _{\;\;\;\;\nu}=-\gamma^{\mu \nu} _{\;\;\;\;\nu}
=-\gamma^{\mu \;\;\;\nu}_{\;\;\nu}
=\gamma^{\nu \;\;\;\nu}_{\;\;\mu}
=\gamma_{\nu}^{\;\;\mu \nu}=-\gamma_{\mu}^{\;\;\nu \nu}
=n_{\mu}\frac{\mathcal{D}'}{\mathcal{D}^3}
\label{gam-nm-n}
\end{eqnarray}
can be obtained.

The contravariant nonzero components of $\gamma^{ikl}$ are
\begin{eqnarray}
\gamma^{\mu00}=-\gamma^{0\mu 0}=-n_{\mu}\frac{\mathcal{D}\mathcal{N}'-\mathcal{N}\mathcal{D}' }{\mathcal{D} \mathcal{N}^3},
\label{gam-up-m00}\\
\gamma^{\nu\mu\nu}=-\gamma^{\mu\nu\nu}=-n_{\mu}\frac{\mathcal{D}'}{\mathcal{D}^5}.
\label{gam-up-mnn}
\end{eqnarray}  
  
Using Eqs.~$(\ref{gam-0_m0})$ and $(\ref{gam-n_mn})$ we get for $\Phi_{m}$ defined by Eq.~$(\ref{def-phiq})$
the result
\begin{eqnarray}
\Phi_{0}=0,
\label{phi-0}\\
\Phi_{\mu}=-n_{\mu}
\Bigl \{\frac{\mathcal{N}'}{\mathcal{N}}+\frac{\mathcal{D}'}{\mathcal{D}}\Bigr \}.
\label{phi-mu}
\end{eqnarray}
The contravariant components of the vector $\Phi$ follow from the metric tensor $g^{mn}$ defined by Eq.~$(\ref{ds2-xyz})$ and 
from Eqs.~$(\ref{phi-0},\ref{phi-mu})$
\begin{eqnarray}
\Phi^{0}=0,
\label{phi-0-up}\\   
\Phi^{\mu}=\frac{n_{\mu}}{\mathcal{D}^2}
\Bigl \{\frac{\mathcal{N}'}{\mathcal{N}}+\frac{\mathcal{D}'}{\mathcal{D}}\Bigr \}.
\label{phi-m-up}
\end{eqnarray}    
\subsection{Solution of the Lagrange equations\\}
\label{appendix-4}

 According to Eqs.~$(\ref{rn-tetr})$ there are two nonzero independent components of the tetrads for the RN
 solution, namely
$h^{(0)}_0$ and $h^{(1)}_1=h^{(2)}_2=h^{(3)}_3$, the latter  will be denoted from here on $h^{(\mu)}_{\mu}$. 
 Substituting into Eq.~$(\ref{lagr-eq})$ $c=p=0$ and $q=\nu=1,\;2,\;3$ ($q \neq 0$ since all field variables are time independent) we get for the  right-hand side of  
Eq.~$(\ref{lagr-eq})$ with the help of Eq.~$(\ref{righ-lagr})$
\begin{eqnarray}
\nonumber
\frac{\partial}{\partial \rho^{\nu}} \Bigl [
\frac{\partial \mathcal{L}_{tot}}{\partial h^{(0)}_{0,\nu}}\Bigr ] 
=\frac{\partial}{\partial \rho^{\nu}} \Bigl [
\frac{|h|}{\kappa} \Bigl ( h_{(0)}^0\gamma^{\nu 0}_{\;\;\;0}+h_{(0)}^0 \Phi^{\nu}
 -h_{(0)}^{\nu}\Phi^{0}\Bigr )\Bigr ]=\\ 
\nonumber
=\frac{\partial}{\partial \rho^{\nu}} \Bigl \{ \frac{\mathcal{N}\mathcal{D}^2 }{\kappa} \Bigl (\frac{\mathcal{D}}{\mathcal{N}}\Bigr )
\Bigl [-\frac{n_{\nu}}{\mathcal{N}\mathcal{D}^3}\Bigl (\mathcal{D}\mathcal{N}'-\mathcal{N}\mathcal{D}'\Bigr )+ \\
+\frac{n_{\nu}}{\mathcal{D}^2} \Bigl ( \frac{\mathcal{N}'}{\mathcal{N}}+\frac{\mathcal{D}'}{\mathcal{D}} \Bigr ) \Bigr ] \Bigr \} 
=\frac{2}{\kappa}\frac{\partial}{\partial \rho^{\nu}}[\mathcal{D}'n_{\nu}].
\label{right-lagr-h00n}
\end{eqnarray}
In the above chain of equations, it is taken into account that 
$\Phi^{0}=0$ according to Eq.~$(\ref{phi-0-up})$ and formula $(\ref{h-up-0-down-0})$ 
for the $h_{(0)}^i$ components. 
Equations $(\ref{det-h-nd})$ for $|h|$, $(\ref{gam-m0-0})$ for $\gamma^{\nu 0}_{\;\;\;\;0}$, 
and $(\ref{phi-m-up})$ for $\Phi^{\nu}$ are also used. Using 
Eq.~$(\ref{n_lam})$ for $n_{\nu}$
and Eq.~$(\ref{d-prime})$ for $\mathcal{D}'$ one gets
\begin{eqnarray}
\frac{\partial}{\partial \rho^{\nu}} \Bigl [
\frac{\partial \mathcal{L}_{tot}}{\partial h^{(0)}_{0,\nu}}\Bigr ]=-\frac{r^2_0}{\kappa \rho^4}-\frac{4 \pi r_g}{\kappa} 
\delta({\bf R})
=-\frac{r^2_0}{\kappa \rho^4}.
\label{right-lagr-h00n-fin}
\end{eqnarray}
Since  $\rho\geq \rho_{min}>0$ for the uniform coordinates
the term with the three-dimensional Dirac delta-function $\delta({\bf R})$
in Eq.~$(\ref{right-lagr-h00n-fin})$ is zero.

In order to obtain the left-hand side of Eq.~$(\ref{lagr-eq})$, we substitute into Eq.~$(\ref{fin-ltot-var-hap})$ $p=c=0$. This leads to the formula
\begin{eqnarray}   
\nonumber
\frac{\partial \mathcal{L}_{tot}}{\partial h^{(0)}_{0}}=\sum_{l=0}^3\frac{|h|}{4 \pi}F^{0l}F_{0l}h^0_{(0)}+h^0_{(0)}\mathcal{L}_{tot} +\\
\nonumber
+\frac{|h|}{\kappa}h^0_{(0)} \Bigl \{ \sum_{m,l=0}^3\gamma^{0ml}\Bigl ( \gamma_{ml0}+\gamma_{l0m}\Bigr ) +\\
+\sum_{\mu=1}^3\Bigl (\gamma^{0\mu}_{\;\;\;\;0}+\gamma_0^{\;\;0\mu}  \Bigr ) \Phi_{\mu} \Bigr \}.
\label{ltot-var-h00}
\end{eqnarray}  
Here, it is taken into consideration that according to Eq.~$(\ref{h-up-0-down-0})$ if $c=0$  the only nonzero component 
of $h_{(c)}^n$ is  $h_{(0)}^0$   and vice versa:  for $n=0$ $h_{(c)}^n \neq 0$ only for $c=0$. 
It is taken into account also that $\Phi^0=\Phi_0=0$ according to Eqs.~$(\ref{phi-0})$ and $(\ref{phi-0-up})$, therefore we may replace $\Phi_{m}$ by  $\Phi_{\mu}$. 

When the 
variable $\rho$ is used instead of $r$ the nonzero component $F_{01}$ is transformed to 
$\hat{F}_{01}=\frac{d r}{d \rho}F_{01}=\mathcal{N}F_{01}$ according to Eq.~$(\ref{def-N})$. The coordinate transformation from $\rho$, $\theta$, 
$\varphi$ to $\rho^{1},\rho^{2},\rho^{3}$
 causes the electromagnetic tensor transformation $\tilde{F}_{0\lambda}=\frac{d \rho}{d 
\rho^{\lambda}}\hat{F}_{01}$ with $\frac{d \rho}{d \rho^{\lambda}}=n_{\lambda}$. Therefore the electromagnetic tensor components for the
uniform coordinates $\rho^k$ are $\tilde{F}_{0\lambda}=n_{\lambda} \mathcal{N}F_{01}$.
Using Eq.~$(\ref{sol-F01})$ for $F_{01}$ and relation~$(\ref{r-rho})$ 
we came to the final formula for the nonzero
components of $\tilde{F}_{ik}$ 
\begin{eqnarray}
F_{0\lambda} =-F_{\lambda 0} ={\bf E}_{\lambda}=\frac{e}{r^2}\mathcal{N}n_{\lambda}=\frac{e \mathcal{N}}{\rho^2\mathcal{D}^2} n_{\lambda}.
\label{f0l-uni}
\end{eqnarray}
In Eq.~$(\ref{f0l-uni})$ and hereafter, more simple notations $F_{ik}$ are used instead of $\tilde{F}_{ik}$ in the uniform coordinate system.
Since $F_{0l}$ is zero for $l=0$ we should write
\begin{eqnarray}
\nonumber
\sum_{l=0}^3F^{0l}F_{0l}=\sum_{\lambda=1}^3g^{00}g^{\lambda \lambda}[F_{0\lambda}]^2 = \\
 =\frac{\mathcal{D}^2}{\mathcal{N}^2}\Bigl( -\frac{1}{\mathcal{D}^2}\Bigr )\sum_{\lambda=1}^3\Bigl [\frac{e n_{\lambda}}{\rho^2\mathcal{D}^2} 
\mathcal{N}\Bigr ]^2=-\frac{e^2}{\rho^4 \mathcal{D}^4}.
\label{aux-fik-fik}
\end{eqnarray}
In transformation of Eq.~$(\ref{aux-fik-fik})$, the obvious relation
\begin{eqnarray}
\sum_{\lambda=1}^3 n_{\lambda}^2=1,
\label{n2=1}
\end{eqnarray}
 and Eq.~$(\ref{ds2-xyz})$ for $g^{jj}=1/ g_{jj}$ are taken into account. If  
Eq.~(\ref{det-h-nd})   for $|h|$, Eq.~(\ref{h-up-0-down-0}) for $h_0^{(0)}$, 
 Eq.~$(\ref{e-re-kap})$ to express $e^2$ through $r_e^2$, and Eq.~$(\ref{aux-fik-fik})$ are used  this leads to the formula
\begin{eqnarray}
h^0_{(0)}\frac{|h|}{4 \pi}\sum_{l=0}^3F^{0l}F_{0l}=-\frac{2r_e^2}{\kappa \rho^4\mathcal{D}}.
\label{sum-fik-2}         
\end{eqnarray}  

According to Eq.~$(\ref{gam-up-m00})$ the contravariant tensor $\gamma^{0ml}$ is nonzero if two indexes are zero. Due to the antisymmetry with respect
to the first and second indexes, $m$ cannot be zero in $\gamma^{0ml}$. Hence $l=0$ in $\gamma^{0ml}$. Due to the same reason $\gamma_{l0m}=0$ if $l=0$. 
Therefore the sum in Eq.~$(\ref{ltot-var-h00})$ can be simplified 
\begin{eqnarray} 
\sum_{m,l=0}^3\gamma^{0ml}\Bigl ( \gamma_{ml0}+\gamma_{l0m}\Bigr )
=\sum_{\mu=1}^3\gamma^{0\mu0}\gamma_{\mu00}.
\label{sum-gamma-1}
\end{eqnarray}
Finally, using Eq.~$(\ref{sum-fik-2})$, formula $(\ref{ltot-nd})$ for $\mathcal{L}_{tot}$, Eq.~$(\ref{h-up-0-down-0})$ for $h^0_{(0)}$, 
substituting Eq.~$(\ref{sum-gamma-1})$ into basic Eq.~$(\ref{ltot-var-h00})$, expressing the nonzero components of the tensor $\gamma$  using
Eqs.~(\ref{gam-m00},  \ref{gam-m0-0},  \ref{gam-up-m00})  and  the four-vector components $\Phi_{\mu}$ with the help of  
$(\ref{phi-mu})$ we get
\begin{eqnarray}
\frac{\partial \mathcal{L}_{tot}}{\partial h^{(0)}_{0}}=-\frac{r_0^2}{\kappa \rho^4}.
\label{fin-ltot-h00}
\end{eqnarray}
In order to obtain the simple formula $(\ref{fin-ltot-h00})$, we use Eqs.~(\ref{def-D},  \ref{def-N},  \ref{n-prime},  \ref{d-prime})
respectively for $\mathcal{D}$, $\mathcal{N}$, $\mathcal{N}'$, $\mathcal{D}'$
taking also into account Eq.~$(\ref{r0})$. 
A comparison of Eqs.~$(\ref{right-lagr-h00n-fin})$  and  $(\ref{fin-ltot-h00})$
with the master equation 
$(\ref{lagr-eq})$ shows that Lagrange equation $(\ref{lagr-eq})$ for $h^{(0)}_0$ is valid. 

Equation $(\ref{lagr-eq})$ for $c=0$ and $p=\nu \neq 0$ is satisfied also, though both the right-hand side and left-hand 
side are zero. Indeed, remembering that $q \neq 0$ in Eq.~$(\ref{righ-lagr})$
since all functions are time independent, hence
$q=\lambda$ with $\lambda=1,\;2,\;3$, we have for this case from Eq.~$(\ref{righ-lagr})$
\begin{eqnarray}
\nonumber
\frac{\partial}{\partial \rho^{\lambda}} \Bigl [
\frac{\partial \mathcal{L}_{tot}}{\partial h^{(0)}_{\nu, \lambda}} \Bigr ]=
\frac{\partial}{\partial \rho^{\lambda}}\Bigl \{ \frac{|h|}{\kappa}\Bigl ( h_{(0)}^{i}\gamma^{\lambda \nu}_{\;\;\;i} +\\
+h_{(0)}^{\nu}\Phi^{\lambda}-h_{(0)}^{\lambda}\Phi^{\nu}\Bigr )\Bigr \}=0. 
\label{right-h0nu}
\end{eqnarray}
We take into consideration that $h_{(0)}^{\nu}=h_{(0)}^{\lambda}=0$ in Eq.~$(\ref{right-h0nu})$ in the uniform coordinates and also 
that $h_{(0)}^{i} \neq 0$ only if $i=0$. The simplest consequence of the last relation is the following  equality: $\gamma^{\lambda 
\nu}_{\;\;\;\;i}=\gamma^{\lambda 
\nu}_{\;\;\;\;0}=0$ valid according to Eqs.~$(\ref{gam-m0-0})$ and $(\ref{gam-nm-n})$ showing nonzero tensor components of
$\gamma^{ik}_{\;\;\;l}$. This proves Eq.~$(\ref{right-h0nu})$.

For the calculation of the left-hand side of
Eq.~$(\ref{lagr-eq})$, we substitute into Eq.~$(\ref{fin-ltot-var-hap})$ $c=0$, $p=\nu$ that gives
\begin{eqnarray}
\nonumber   
\frac{\partial \mathcal{L}_{tot}}{\partial h^{(0)}_{\nu}}=\frac{|h|}{4 \pi}
F^{\nu 0}F_{i0}h^{i}_{(0)}+h^{\nu}_{(0)}\mathcal{L}_{tot}+ \\
\nonumber
+\frac{|h|}{\kappa}h^i_{(0)} \Bigl \{ \sum_{m,l=0}^3\gamma^{\nu ml}\Bigl ( \gamma_{mli}+\gamma_{lim}\Bigr ) + \\
+\sum_{\mu=1}^3\Bigl (\gamma^{\nu\mu}_{\;\;\;\;i}+\gamma_i^{\;\;\nu\mu}  \Bigr ) \Phi_{\mu} +\Phi_i \Phi^{\nu}\Bigr \}.
\label{ltot-var-h0nu}
\end{eqnarray}   
Since $h^i_{(0)} \neq 0$ only for $i=0$, then the first  term in 
Eq.~$(\ref{ltot-var-h0nu})$ is zero as $F_{00}=0$, while the second term vanishes as $h^{\nu}_{(0)} = 0$. The tensor components 
$\gamma^{\nu\mu}_{\;\;\;\;i}=\gamma_i^{\;\;\nu\mu}=0$ at 
$i=0$ since $\gamma^{nm}_{\;\;\;\;i} \neq 0$ when even number of indexes are equal to zero, and also $\Phi_i=0$ at $i=0$ 
according to Eq.~$(\ref{phi-0})$. Therefore two last terms in the curl brackets in Eq.~$(\ref{ltot-var-h0nu})$ are zero. In the sum over $m$ and $l$ 
in Eq.~$(\ref{ltot-var-h0nu})$, the tensor 
$\gamma^{\nu m l}$ is nonzero according to Eqs.~$(\ref{gam-up-m00})$ and $(\ref{gam-up-mnn})$ 
for two cases: $m=l=0$ or both  $m$ and $l$ are nonzero. For the former case and 
for $i=0$ $\gamma_{mli}=\gamma_{lim}=\gamma_{000}=0$, while for the latter case $\gamma_{m l0}=\gamma_{l 0 m}=0$ 
since they are absent in 
Eqs.~$(\ref{gam-m00})$ and $(\ref{gam-nmn})$ for nonzero 
$\gamma_{ikl}$. As a result we get
\begin{eqnarray}
\frac{\partial \mathcal{L}_{tot}}{\partial h^{(0)}_{\nu}}=0,
\label{ltot-var-h0nu-zero}
\end{eqnarray}  
that proves the validity of Lagrange equations $(\ref{lagr-eq})$ for this case.

For $p=c=\mu$ and $q=\lambda$   Eq.~$(\ref{righ-lagr})$ looks like
\begin{eqnarray}
\nonumber
\sum_{\lambda=1}^3\frac{\partial}{\partial \rho^{\lambda}} \Bigl [
\frac{\partial \mathcal{L}_{tot}}{\partial h^{(\mu)}_{\mu,\lambda}}\Bigr ]= \\
=\sum_{\lambda=1}^3 \frac{\partial}{\partial \rho^{\lambda}} \Bigl [
\frac{|h|}{\kappa} \Bigl ( h_{(\mu)}^{\mu}\gamma^{\lambda \mu}_{\;\;\;\;\mu}+h_{(\mu)}^{\mu} \Phi^{\lambda}
 -h_{(\mu)}^{\lambda}\Phi^{\mu}\Bigr )\Bigr ].
\label{dltot-dhmm}
\end{eqnarray}
Expressing $h_{(\mu)}^{\lambda}$ with the help of Eq.~$(\ref{hlm-obv})$,
using Eq.~$(\ref{phi-m-up})$ for $\Phi^{\lambda}$ and Eq.~$(\ref{det-h-nd})$ for the determinant $|h|$, Eq.~$(\ref{dltot-dhmm})$ is 
transformed to the following:
\begin{eqnarray}
\nonumber
\sum_{\lambda=1}^3\frac{\partial}{\partial \rho^{\lambda}} \Bigl [
\frac{\partial \mathcal{L}_{tot}}{\partial h^{(\mu)}_{\mu,\lambda}}\Bigr ] 
= -\sum_{\lambda=1}^3\frac{\partial}{\partial \rho^{\lambda}} \Bigl \{
\frac{\mathcal{N}\mathcal{D}^2}{\kappa} \Bigl [ \frac{\gamma^{\lambda \mu}_{\;\;\;\;\mu}}{\mathcal{D}}+  \\
+\frac{n_{\lambda}-n_{\mu}\delta^{\lambda}_{\mu}} 
{\mathcal{D}^3}\Bigl (\frac{\mathcal{N}'}{\mathcal{N}}+\frac{\mathcal{D}'}{\mathcal{D}} \Bigr )\Bigr ]\Bigr \}.
\label{dltot-dhmm-1}
\end{eqnarray}
Since both $\gamma^{\lambda \mu}_{\;\;\;\;\mu}=0$ and $n_{\lambda}-n_{\mu}\delta^{\lambda}_{\mu}=0$ at $\lambda =\mu$, then 
Eq.~$(\ref{dltot-dhmm-1})$
can be rewritten in the form
\begin{eqnarray}
\nonumber
\sum_{\lambda=1}^3\frac{\partial}{\partial \rho^{\lambda}} \Bigl [
\frac{\partial \mathcal{L}_{tot}}{\partial h^{(\mu)}_{\mu,\lambda}}\Bigr ] 
= \sum_{\lambda \neq \mu}\frac{\partial}{\partial \rho^{\lambda}} \Bigl \{
\frac{\mathcal{N}\mathcal{D}^2}{\kappa} \Bigl [ \frac{n_{\lambda}\mathcal{D}' }{\mathcal{D}^4}- \\
-\frac{n_{\lambda}}
{\mathcal{D}^3}\Bigl (\frac{\mathcal{N}'}{\mathcal{N}}+\frac{\mathcal{D}'}{\mathcal{D}} \Bigr )\Bigr ]\Bigr \}
=-\sum_{\lambda \neq \mu}\frac{\partial}{\partial \rho^{\lambda}} \Bigl \{
\frac{\mathcal{N}' n_{\lambda}}{\kappa \mathcal{D}}\Bigr \}.
\label{dltot-dhmm-2}
\end{eqnarray} 
Here, Eq.~$(\ref{gam-nm-n})$ for $\gamma^{\lambda \mu}_{\;\;\;\;\mu}$
 is taken into account.
Finally,  using Eq.~$(\ref{def-D})$ for $\mathcal{D}$ and 
 Eq.~$(\ref{n-prime})$ for $\mathcal{N}'$ we obtain the formula for the 
right-hand side of Eq.~$(\ref{lagr-eq})$ for $p=c=\mu$
\begin{eqnarray}
\nonumber
\sum_{\lambda=1}^3\frac{\partial}{\partial \rho^{\lambda}} \Bigl [
\frac{\partial \mathcal{L}_{tot}}{\partial h^{(\mu)}_{\mu,\lambda}}\Bigr ] = \\
=-\frac{r_0^2}{\kappa \rho^4 \mathcal{D}^2}\Bigl [ 1+\frac{r_g}{4\rho} 
-2n_{\mu}^2\Bigl ( 1+\frac{3r_g}{8\rho}-\frac{r_0^2}{8 \rho^2}\Bigr ) \Bigr ].
\label{dltot-dhmm-fin}
\end{eqnarray}

For the left-hand side of Eq.~$(\ref{lagr-eq})$ for $p=c=\mu$ we have according to Eq.~$(\ref{fin-ltot-var-hap})$
\begin{eqnarray}
\nonumber
\frac{\partial \mathcal{L}_{tot}}{\partial h^{(\mu)}_{\mu}}=\frac{|h|}{4 \pi}\sum_{l=0}^3
F^{\mu l}F_{\mu l}h^{\mu}_{(\mu)}+h^{\mu}_{(\mu)}\mathcal{L}_{tot}+ \\
\nonumber
+\frac{|h|}{\kappa}h^{\mu}_{(\mu)} \Bigl \{ \sum_{m=0}^3 \sum_{l=0}^3\gamma^{\mu ml}\Bigl ( \gamma_{ml \mu}+\gamma_{l \mu m}\Bigr ) +\\
+\sum_{m=0}^3  \Bigl (\gamma^{\mu m}_{\;\;\;\;\;\mu}+\gamma_{\mu}^{\;\;\mu m}  \Bigr ) \Phi_m +\Phi_{\mu} \Phi^{\mu} \Bigr \}.
\label{ltot-var-hmm}
\end{eqnarray}
Since $F_{\mu l}$ is nonzero only if $l=0$, then the sum over $l$ in the first term in the right-hand side of Eq.~$(\ref{ltot-var-hmm})$ contains one term 
with 
$l=0$. Using 
Eq.~$(\ref{f0l-uni})$ for $F_{\mu 0}$,  formulas for $g^{00}$, $g^{\mu\mu}$  which follow from Eq.~$(\ref{ds2-xyz})$, Eq.~$(\ref{det-h-nd})$ 
for $|h|$, and  
Eq.~$(\ref{hlm-obv})$ for $h^{\mu}_{(\mu)}$ we get
\begin{eqnarray}
\frac{|h|}{4 \pi}\sum_{l=0}^3F^{\mu l}F_{\mu l}h^{\mu}_{(\mu)}=\frac{|h|}{4 \pi}g^{00}g^{\mu\mu}(F_{\mu 0})^2h^{\mu}_{(\mu)}=
n_{\mu}^2\frac{2 r_e^2}{\kappa \rho^4}\frac{\mathcal{N}}{\mathcal{D}^3}
\label{aux-01}
\end{eqnarray}
in an analogous way as Eq.~$(\ref{sum-fik-2})$ was obtained. 

Remembering property of $\gamma^{\mu ml}$ presented by Eqs.~$(\ref{gam-up-m00})$ and 
$(\ref{gam-up-mnn})$ we conclude that the case $m=l=0$ gives one term in the sum over $m$ and $l$ in the curl brackets in 
Eq.~$(\ref{ltot-var-hmm})$ 
which is equal to $\gamma^{\mu 00}\gamma_{0\mu 0}$ since $\gamma_{00\mu}=0$ due the antisymmetry of $\gamma_{ik\mu}$ with respect to $i$ and $k$. Other 
nonzero terms in the same sum over $m$ and $l$ correspond to the 
cases when $m=l=\nu$ with $\nu \neq \mu$,  or $m=\nu$, $l=\mu$ with $\nu \neq \mu$ in any case. The net result is
\begin{eqnarray}
\nonumber
\sum_{m=0}^3 \sum_{l=0}^3\gamma^{\mu ml}\Bigl ( \gamma_{ml \mu}+\gamma_{l \mu m}\Bigr )=
\gamma^{\mu 00}\gamma_{0\mu 0}+\\
+\sum_{\nu \neq \mu}\Bigl (\gamma^{\mu \nu \nu}\gamma_{\nu \mu \nu} 
+\gamma^{\mu \nu \mu}\gamma_{\nu \mu \mu}\Bigr ).  
\label{aux-02}
\end{eqnarray}
Using Eqs.~$(\ref{gam-m00})$ and $(\ref{gam-up-m00})$ we have
\begin{eqnarray}
\gamma^{\mu 00}\gamma_{0\mu 0}=\frac{n^2_{\mu}}{\mathcal{N}^2\mathcal{D}^4}\{\mathcal{D}\mathcal{N}'-\mathcal{N}\mathcal{D}'\}^2.
\label{aux-03}
\end{eqnarray}
The first sum over $\nu$ in Eq.~$(\ref{aux-02})$ is also easily calculated
\begin{eqnarray}
\sum_{\nu \neq \mu}\gamma^{\mu \nu \nu}\gamma_{\nu \mu \nu}=\sum_{\nu \neq \mu}n^2_{\mu}\frac{(\mathcal{D}')^2}{\mathcal{D}^4}
=2n^2_{\mu}\frac{(\mathcal{D}')^2}{\mathcal{D}^4}
\label{aux-04}
\end{eqnarray}
if Eqs.~ $(\ref{gam-nmn})$ and $(\ref{gam-up-mnn})$  for  $\gamma_{\nu \mu \nu}$ and  $\gamma^{\mu \nu \nu}$
 are used, respectively.  For transformation of the second sum in Eq.~$(\ref{aux-02})$, one writes
\begin{eqnarray}
\sum_{\nu \neq \mu}\gamma^{\mu \nu \mu}\gamma_{\nu \mu \mu}=\sum_{\nu \neq \mu}\frac{(\mathcal{D}')^2}{\mathcal{D}^4}n^2_{\nu} 
=(1-n^2_{\mu})\frac{(\mathcal{D}')^2}{\mathcal{D}^4}.
\label{aux-05}
\end{eqnarray}
Here, the obvious relation for the unit three-vector $n_{\nu}$
\begin{eqnarray}
\sum_{\nu \neq \mu}n^2_{\nu}=1-n^2_{\mu}
\label{obv-02}
\end{eqnarray}
is used in addition to formulas $(\ref{gam-nmn})$ and $(\ref{gam-up-mnn})$.

Since in the last sum over $m$ in the curl brackets in Eq.~$(\ref{ltot-var-hmm})$ the term with $m=0$ is zero according 
to Eq.~$(\ref{phi-0})$ and $\gamma_{\mu}^{\;\;\mu m}=0$ due to the antisymmetry property, 
then using Eqs.~$(\ref{gam-nm-n})$ and $(\ref{phi-mu})$ we get
\begin{eqnarray}
\nonumber
\sum_{\nu =1}^3\gamma^{\mu \nu}_{\;\;\;\;\mu} \Phi_{\nu}=-\sum_{\nu \neq \mu}n_{\nu}^2 \frac{\mathcal{D}'}{\mathcal{D}^3}
\Bigl ( \frac{\mathcal{N}'}{\mathcal{N}}+\frac{\mathcal{D}'}{\mathcal{D}} \Bigr )= \\
=(n_{\mu}^2-1)\frac{\mathcal{D}'}{\mathcal{D}^3}\Bigl ( \frac{\mathcal{N}'}{\mathcal{N}}+\frac{\mathcal{D}'}{\mathcal{D}} \Bigr ).
\label{aux-06}
\end{eqnarray}
Equation $(\ref{obv-02})$ is again used to transform Eq.~$(\ref{aux-06})$.
 
Finally, making use of Eq.~$(\ref{det-h-nd})$ for the determinant $|h|$,
Eq.~$(\ref{hlm-obv})$ for $h^{\mu}_{(\mu)}$,  Eq.~$(\ref{ltot-nd})$ for $\mathcal{L}_{tot}$,
substituting Eqs.~(\ref{aux-01}--\ref{aux-05}) and $(\ref{aux-06})$  into formula $(\ref{ltot-var-hmm})$, using also 
Eqs.~$(\ref{phi-mu})$ and $(\ref{phi-m-up})$ to obtain $\Phi_{\mu} \Phi^{\mu}$, we get  
\begin{eqnarray}
\frac{\partial \mathcal{L}_{tot}}{\partial h^{(\mu)}_{\mu}}
=-\frac{r_0^2}{\kappa \rho^4 \mathcal{D}^2}\Bigl [ 1+\frac{r_g}{4\rho}
-2n_{\mu}^2\Bigl ( 1+\frac{3r_g}{8\rho}-\frac{r_0^2}{8 \rho^2}\Bigr ) \Bigr ].
\label{fin-lhs-hmm}
\end{eqnarray}
A comparison of Eqs.~$(\ref{fin-lhs-hmm})$ with $(\ref{dltot-dhmm-fin})$ shows that Lagrange equation $(\ref{lagr-eq})$  for 
$p=c=\mu$ is satisfied.

Though the tetrad components $h^{(\mu)}_{\nu}$ for $\mu \neq \nu$ are zero, the Lagrange equation for this case is nontrivial, and it 
will be shown that it is satisfied for the tetrad given by Eqs.~$(\ref{rn-tetr})$. Substituting in Eq.~$(\ref{righ-lagr})$
$c=\mu$, $p=\nu$, $q=\lambda$ and remembering that $h^i_{(c)}$ is nonzero only if $i=c$ we write
\begin{eqnarray}
\nonumber
\sum_{\lambda=1}^3\frac{\partial}{\partial \rho^{\lambda}} \Bigl [
\frac{\partial \mathcal{L}_{tot}}{\partial h^{(\mu)}_{\nu, \lambda}} \Bigr ]=
\sum_{\lambda=1}^3\frac{\partial}{\partial \rho^{\lambda}}\Bigl \{ \frac{|h|}{\kappa}\Bigl ( h_{(\mu)}^{\mu}
\gamma^{\lambda \nu}_{\;\;\;\;\mu} +\\
+\;h_{(\mu)}^{\nu}\Phi^{\lambda}-h_{(\mu)}^{\lambda}\Phi^{\nu}\Bigr )\Bigr \}.
\label{right-hmunu}
\end{eqnarray}  
Since $\gamma^{\lambda \nu}_{\;\;\;\;\mu}=0$ if all indexes are different from each other 
or $\lambda =\nu$ due to the antisymmetry with respect to  these indexes, hence 
$\gamma^{\lambda \nu}_{\;\;\;\;\mu}\neq 0$ for $\lambda = \mu$ only. The second term in the brackets in 
Eq.~$(\ref{right-hmunu})$ is zero  as $h_{(\mu)}^{\nu}=0$ for $\mu \neq \nu$,
while the third term is nonzero only for $\lambda = \mu$. Therefore Eq.~$(\ref{right-hmunu})$ can be transformed to 
\begin{eqnarray}
\sum_{\lambda=1}^3\frac{\partial}{\partial \rho^{\lambda}} \Bigl [
\frac{\partial \mathcal{L}_{tot}}{\partial h^{(\mu)}_{\nu, \lambda}} \Bigr ]=
\frac{\partial}{\partial \rho^{\mu}}\Bigl \{ \frac{|h|}{\kappa}h_{(\mu)}^{\mu} \Bigl 
(\gamma^{\mu \nu}_{\;\;\;\;\mu} -\Phi^{\nu}\Bigr )\Bigr \}.
\label{right-h0nu-1}
\end{eqnarray}
Using Eqs.~(\ref{det-h-nd},  \ref{hlm-obv},  \ref{gam-nm-n}), and $(\ref{phi-m-up})$ respectively for  
$|h|$, $h_{(\mu)}^{\mu}$,  $\gamma^{\mu \nu}_{\;\;\;\;\mu}$, and $\Phi^{\nu}$ we get the final result
\begin{eqnarray}
\nonumber
\sum_{\lambda=1}^3\frac{\partial}{\partial \rho^{\lambda}} \Bigl [
\frac{\partial \mathcal{L}_{tot}}{\partial h^{(\mu)}_{\nu, \lambda}} \Bigr ]=
-\frac{\partial}{\partial \rho^{\mu}}\Bigl \{ \frac{n_{\nu}\mathcal{N}'}{\kappa \mathcal{D}}
\Bigr \} =\\
=\frac{2r_0^2n_{\mu}n_{\nu}}{\kappa \mathcal{D}^2 \rho^4}\Bigl \{
1+\frac{3 r_g}{8\rho}-\frac{r_0^2}{8\rho^2}\Bigr \}.
\label{right-h0nu-1a}
\end{eqnarray}
Here, Eqs.~$(\ref{def-D})$ and  $(\ref{n-prime})$
respectively for $\mathcal{D}$ and  $\mathcal{N}'$    are taken into account.

For 
$\partial \mathcal{L}_{tot}/\partial h^{(\mu)}_{\nu}$ we have  from 
Eq.~$(\ref{fin-ltot-var-hap})$
\begin{eqnarray}
\nonumber
\frac{\partial \mathcal{L}_{tot}}{\partial h^{(\mu)}_{\nu}}=\frac{|h|}{4 \pi}F^{\nu 0}F_{\mu 0}h^{\mu}_{(\mu)}+ \\
\nonumber
+\frac{|h|}{\kappa}h^{\mu}_{(\mu)} \Bigl \{ \sum_{m,l=0}^3\gamma^{\nu m l}\Bigl ( \gamma_{m l \mu}+\gamma_{l \mu m}\Bigr )+\\
+\sum_{m=0}^3\Bigl (\gamma^{\nu m}_{\;\;\;\;\mu}+\gamma_{\mu}^{\;\;\nu m}  \Bigr ) \Phi_{m} +\Phi_{\mu} \Phi^{\nu} \Bigr \}.
\label{fin-ltot-var-hmunu}
\end{eqnarray}
It is taken into account in Eq.~$(\ref{fin-ltot-var-hmunu})$ that $F^{\nu l}F_{\mu l}$ is nonzero for $l=0$ only, and $h_{(\mu)}^{\nu} \neq 0$
only if $\mu=\nu$. The term $F^{\nu 0}F_{\mu 0}/(4 \pi)$ is transformed as
\begin{eqnarray}
\frac{F^{\nu 0}F_{\mu 0}}{4 \pi}=\frac{1}{4 \pi}g^{\nu\nu}g^{00} F_{\nu 0}F_{\mu 0}=-\frac{2r_e^2}{\kappa}\frac{n_{\nu}n_{\mu}}{\rho^4D^4}
\label{tran-el}
\end{eqnarray}
with the help of  Eq.~$(\ref{e-re-kap})$ and Eq.~$(\ref{f0l-uni})$ for $F_{\nu 0}$ and $F_{\mu 0}$. Formula~$(\ref{ds2-xyz})$ is also used for $g^{\nu\nu}=1/g_{\nu\nu}$ and 
$g^{00}=1/g_{00}$ in Eq.~$(\ref{tran-el})$.

In the sum over $m$ and $l$ in the curl brackets of Eq.~$(\ref{fin-ltot-var-hmunu})$, there is the 
nonzero term with $m=l=0$ and the 
term with  $m=l=\lambda$, $\lambda \neq \nu$, $\lambda \neq \mu$ according to the properties of $\gamma^{jkl}$ and $\gamma_{nmq}$, therefore 
\begin{eqnarray}
\sum_{m,l=0}^3\gamma^{\nu m l}\Bigl ( \gamma_{m l \mu}+\gamma_{l \mu m}\Bigr )=\gamma^{\nu 00}\gamma_{0 \mu 0}+
\gamma^{\nu \lambda \lambda}\gamma_{\lambda \mu \lambda}.
\label{sum-ml}
\end{eqnarray}
We remember that $\gamma_{00 \mu}=\gamma_{\lambda \lambda \mu}=0$ owing to the antisymmetry of $\gamma_{ikn}$ with respect to $i$ and $k$. 
Note that $\gamma^{\nu m l} \neq 0$ in Eq.~$(\ref{sum-ml})$ for $l=\nu$ and $m=\lambda \neq \nu$ but $\gamma_{m l \mu}+\gamma_{l \mu m}=0$ for all these $m$ and $l$
according to Eq.~$(\ref{gam-nmn})$.

In the last sum 
over $m$ in the right-hand side of Eq.~$(\ref{fin-ltot-var-hmunu})$ the nonzero terms can be only for $m=\mu$ or $m=\nu$. This gives
\begin{eqnarray}
\sum_{m=0}^3\Bigl (\gamma^{\nu m}_{\;\;\;\;\mu}+\gamma_{\mu}^{\;\;\nu m}  \Bigr ) \Phi_{m}
=\gamma^{\nu \mu}_{\;\;\;\;\mu}\Phi_{\mu} 
+\gamma_{\mu}^{\;\;\nu \mu}\Phi_{\mu} +\gamma_{\mu}^{\;\;\nu \nu}\Phi_{\nu}
\label{sum-m-phi}
\end{eqnarray}
since $\gamma^{\nu \nu}_{\;\;\;\;\mu}=0$.
Substituting formulas (\ref{tran-el}--\ref{sum-m-phi}) into Eq.~$(\ref{fin-ltot-var-hmunu})$ and using 
Eqs.~(\ref{det-h-nd},  \ref{hlm-obv}) 
respectively for 
$|h|$, $h_{(\mu)}^{\mu}$, and also Eqs.~(\ref{gam-m00}--\ref{gam-up-mnn}) for
the needed tensor components of $\gamma_{jkl}$, $\gamma^{jkl}$, $\gamma_j^{\;\;kl}$, 
$\gamma^j_{\;\;kl}$
and Eqs.~(\ref{phi-mu}, \ref{phi-m-up}) for the vector components of $\Phi^{\nu}$,  $\Phi_{\mu}$, respectively, we obtain finally
\begin{eqnarray}
\frac{\partial \mathcal{L}_{tot}}{\partial h^{(\mu)}_{\nu}}
=\frac{2r_0^2n_{\mu}n_{\nu}}{\kappa \mathcal{D}^2 \rho^4}\Bigl \{
1+\frac{3 r_g}{8\rho}-\frac{r_0^2}{8\rho^2}\Bigr \}.
\label{ltot-var-hmunu-fin}
\end{eqnarray}
Note that $r^2_e$ from Eq.~$(\ref{tran-el})$ is combined with $r_g^2/4$ from other terms to give $r_0^2$ according to Eq.~$(\ref{r0})$. 
A comparison of Eq.~$(\ref{ltot-var-hmunu-fin})$ with  Eq.~$(\ref{right-h0nu-1a})$ confirms that Lagrange equations $(\ref{lagr-eq})$ for 
$p=\nu$, $c=\mu$, and $\nu \neq \mu$ are fulfilled.

\end{document}